%% file: main.tex
\documentclass[10pt]{iopart}

\expandafter\let\csname equation*\endcsname\relax
\expandafter\let\csname endequation*\endcsname\relax

\usepackage{amsmath}
\usepackage{graphicx}
\usepackage{subcaption}
\usepackage{xcolor}
\usepackage{multicol}
\usepackage{tikz}
\usetikzlibrary{positioning,shapes,shapes.multipart,arrows.meta,calc,fit}
\usepackage{pgfplots}
\usepackage[draft]{todonotes}
\pgfplotsset{compat=1.5.1}
\usepackage[boxed, ruled, lined]{algorithm2e}
\usepackage{booktabs}
\usepackage{rotating}

\usepackage{enumitem}
\usepackage{hyperref}

\pgfplotsset{width=4.5cm,compat=1.5.1}

\input{Mathe-Definitionen}

\begin{document}

\title[Widely Differing Particle Concentrations in MPI]{Simultaneous Imaging of Widely Differing Particle Concentrations in MPI: Problem Statement and Algorithmic Proposal for Improvement}

\author{Marija Boberg, Nadine Gdaniec, Patryk Szwargulski, Franziska Werner, Martin M\"oddel, and Tobias Knopp}

\address{Section for Biomedical Imaging, University Medical Center Hamburg-Eppendorf, 20246 Hamburg, Germany}
\address{Institute for Biomedical Imaging, Hamburg University of Technology, 21073 Hamburg, Germany}
\ead{\mailto{m.boberg@uke.de} and \mailto{t.knopp@uke.de}}

\begin{abstract}
Magnetic Particle Imaging (MPI) is a tomographic imaging technique for determining the spatial distribution of superparamagnetic nanoparticles. Current MPI systems are capable of imaging iron masses over a wide dynamic range of more than four orders of magnitude. In theory, this range could be further increased using adaptive amplifiers, which prevent signal clipping. While this applies to a single sample, the dynamic range is severely limited if several samples with different concentrations or strongly inhomogeneous particle distributions are considered. One scenario that occurs quite frequently in pre-clinical applications is that a highly concentrated tracer bolus in the vascular system ``shadows'' nearby organs with lower effective tracer concentrations. The root cause of the problem is the ill-posedness of the MPI imaging operator, which requires regularization for stable reconstruction. In this work, we introduce a simple two-step algorithm that increases the dynamic range by a factor of four. Furthermore, the algorithm enables spatially adaptive regularization, i.e. highly concentrated signals can be reconstructed with maximum spatial resolution, while low concentrated signals are strongly regularized to prevent noise amplification. 
\end{abstract}

\section{Introduction}
Magnetic particle imaging (MPI) is a tomographic imaging method \cite{knopp2017magnetic} for the quantitative determination of the local concentration of iron-oxide nanoparticles~\cite{Weizenecker2007PhysMedBio,vogel2016first,Graeser2017SR}. Medical applications can be found e.g.~in imaging vascular structures, which enables the detection of stroke \cite{ludewig2017magnetic,Szwargulski2020}, gut bleeding \cite{yu2017magneticB}, cancer \cite{arami2017tomographic}, stenoses \cite{vaalma2017magnetic}, and the presence of cerebral aneurysms \cite{sedlacik2016magnetic}. The very high imaging speed of MPI of up to $46$ volumes respectively $1840$ images per second allows blood flow measurements within a single imaging volume \cite{kaul2018magnetic, Vogel2020}. Furthermore, MPI has proven to be suited for visualizing lung perfusion \cite{zhou2017first}, labeled stem cells \cite{bulte2015quantitative}, and cerebral blood volume \cite{cooley2018rodent}. By coating medical instruments with nanoparticles \cite{haegele2016magnetic} it is in addition possible to perform interventional procedures \cite{salamon2016magnetic,herz2018magnetic,haegele2016multi}.\footnote[0]{This is the Accepted Manuscript version of an article accepted for publication in Physics in Medicine \& Biology. IOP Publishing Ltd is not responsible for any errors or omissions in this version of the manuscript or any version derived from it.  The Version of Record is available online at \url{http://doi.org/10.1088/1361-6560/abf202}.}

MPI is very sensitive and can image a single sample with an iron mass of about \SI{5}{\nano \gram \of {Fe}} in today's pre-clinical imaging devices \cite{Graeser2017SR}. Below a certain concentration level, the measured MPI raw data signal is linear in the sample concentration and the detection limit is technically limited by noise in the receiver electronic \cite{loewa2016concentration}. Many MPI systems are capable of imaging an idealized dynamic range of about four orders of magnitude \cite{Graeser2017SR}. Here, we use the term \textit{idealized dynamic range} to express the ratio between the maximum and minimum concentration that can be sequentially imaged when there is only a single sample with a given iron concentration in the measurement field.

Apart from the technical acquisition of a signal, MPI requires the reconstruction of the measured signal into a tomogram. The reconstruction involves the solution of an ill-posed inverse problem, which is commonly handled by adaptive regularization techniques to obtain stable solutions \cite{erb2018mathematical}. Regular reconstruction schemes such as the Tikhonov regularized least squares method for system matrix reconstructions \cite{Knopp2010PhysMedBio} do not impose further restrictions on the dynamic range and enable using the full quantization range of typical ADCs. Consequently, an image can usually be reconstructed as long as the raw data signal is above the noise floor. 

However, the single sample and single concentration scenario considers an idealized dynamic range and does not consider the general problem of how large the dynamic range in the image space is when there are multiple samples of different concentrations in the measuring field. That this is a severe problem has been discussed by \cite{FrankeHybridMRMPI2016,Herz2017}. We therefore introduce the \textit{effective dynamic range} or just \textit{dynamic range} and define it as the maximum concentration ratio of two samples that can be distinguished in the final image. This scenario often comes into play\textit{ in-vivo} where tracer concentrations can vary over several orders of magnitude. High concentrations are found where a large volume fraction of blood is present, such as in the vascular system. In contrast, lower volume fractions and thus lower concentrations can be found in some organs such as the kidneys or brain. For instance, one can estimate a 70 fold lower particle concentration in the brain compared to the concentration in vessels \cite{Oeff1955}.

\subsection{Organization of the Paper}

This paper is organized as follows. In sections~\ref{Sec:ImagEq} and~\ref{Sec:Reco} we recap the discrete MPI imaging equation and the algebraic image reconstruction approach, respectively. In section~\ref{Sec:ProblemStatement} we outline the problem of a limited dynamic range with the current reconstruction method. We show under which circumstances a highly concentrated image structure can ``shadow'' an adjacent low concentrated structure. Based on these findings we introduce an image reconstruction algorithm to increase the dynamic range in section~\ref{Sec:ContrastEnhanced}. The performance of the proposed algorithm is compared to the common system matrix reconstruction method introduced in section~\ref{Sec:Reco} and the idealized dynamic range using \textit{in-vitro} experiments outlined in section \ref{Section:ExpSetup}. An in depth analysis and discussion of the experimental results is provided in sections~\ref{Sec:Results} and~\ref{Sec:Discussion}.

\section{Imaging Equation} \label{Sec:ImagEq}

MPI uses a combination of static and periodic magnetic fields for the excitation of the magnetic moments in magnetic nanoparticles. The resulting ensemble magnetization encodes the spatial distribution and density of the particles. Its temporal derivative can be picked up as voltage signal $\signatur{u}{[0,T)}{\IR}$ by one or more receive coils, where $T$ is the period length of the excitation pattern. Without loss of generalization we consider a single-channel receiver in this section. Apart from the particle signal the receivers pick up the feed-through of the excitation field, which can be filtered out electronically within the analog signal chain of the receiver~\cite{graeser2013analog}. Therefore, the time signal $u$ is often expanded into a Fourier series with coefficients $\hat{u}_k\in\IC, k \in \IZ$. The relation between the particle concentration $\signatur{c}{\IR^3}{\IR_+}$ and the measured Fourier coefficients $\hat{u}_k$ is linear and can be expressed as
\begin{align} \label{Eq:ImagingEquation}
    \hat{u}_k & = \int_{\Omega} \hat{s}_k(\zb r) c(\zb r) \d\zb r,
\end{align}
where $\Omega\subseteq \IR^3$ is the support of the MPI system function $\signatur{\hat{s}_k}{\IR^3}{\IC}$. Due to the complex particle dynamics, the system function $\hat{s}_k$ is challenging to model physically \cite{Kluth2019} and thus it is usually measured within a calibration procedure. The structure of the MPI system matrix was investigated in more depth in \cite{Rahmer2012TMI}.

By discretization of space one obtains the discrete MPI imaging equation 
\begin{align} \label{Eq:ImagDiscrete}
    \hat{u}_k & \approx \sum_{n=1}^{N} \hat{s}_{k,n} c_n,
\end{align}
where $\hat{s}_{k,n} = w_n \hat{s}_k(\zb r_n)$ with quadrature weights $w_n$, $c_n = c(\zb r_n)$, and $\zb r_n$ are the sampling positions for $n\in I_N := \set{1,\dots,N}$. Due to discrete sampling of the induced voltage $u(t)$ only a finite number of $K$ frequency components are available in practice. The discrete imaging equation \eqref{Eq:ImagDiscrete} can then be written in matrix-vector form as
\begin{align} \label{Eq:ImagDiscreteMV}
    \hat{\zb u} & \approx \zb S \zb c,
\end{align}
where $\hat{\zb u} = \left( \hat{u}_k \right)_{k\in I_K}$ is the measurement vector, $\zb c = \left( c_n \right)_{n\in I_N}$ is the particle concentration vector, and 
\begin{align*}
    \zb S = \left( \hat{s}_{k,n} \right)_{k\in I_K;n\in I_N}
\end{align*}
is the system matrix. In practice, $\hat{\zb u}$ is only approximately known, since the measured signal
\begin{align*} 
    \hat{\zb u}_\meas & = \hat{\zb u} + \bet
\end{align*}
is distorted by a background signal $\bet\in\IC^K$ leading to inconsistencies in the linear system of equations~\eqref{Eq:ImagDiscreteMV}. The background signal $\bet$ is composed of various components such as regular Gaussian noise generated by the electronic components, quantization noise induced by the digitizer, and systematic background signals generated e.g. by thermal scanner drifts \cite{straub2018joint,paysen2020characterization}.

\section{Image Reconstruction} \label{Sec:Reco}

The goal of MPI is to determine the unknown particle concentration vector $\zb c$ from the measured voltage vector $\hat{\zb u}_\meas$ while the system matrix $\zb S$ is considered to be known. The regular way of tackling this problem is to consider the least squares problem
\begin{align} \label{Eq:LS}
    \zb c_\text{LS} &= \underset{\zbs c\in \IR^N_+}{\text{argmin}}\norm{ \zb S \zb c - \hat{\zb u}_\meas }_2^2.
\end{align}
It minimizes the residual of the linear system, which is appropriate if the linear system has full rank and $\bet = \zb 0$. However, in MPI the linear system is ill conditioned, which leads to noise amplification bounded only by the condition number of the MPI system matrix. In practice, this number can be larger than $10^3$. One way to handle this problem is to consider only measurements with a signal-to-noise ratio (SNR) larger than the conditioning of $\zb S$. Since the noise level can hardly be influenced in optimized MPI systems, this implies that one could only measure samples of a very high concentration severely reducing the practical value of MPI.

Fortunately, there are methods to handle the noise amplification of the least squares approach~\eqref{Eq:LS} and reconstruct low SNR data. These techniques are grouped together under the term regularization. The basic idea is to penalize non-plausible solutions additionally to the minimization of the residual $\zb S \zb c - \hat{\zb u}_\meas$. 
\begin{enumerate}
    \item A simple, yet powerful technique is called Tikhonov regularization and uses a penalization term $\norm{ \zb c }_2^2$. It prevents the solution $\zb c$ from  containing large oscillations, which happens typically, if $\zb c$ is fitted to match the noise. Tikhonov regularization can be expressed as
    \begin{align*} 
        \zb c_\text{Reg}^\lambda &= \underset{\zbs c\in \IR^N_+}{\text{argmin}}\norm{ \zb S \zb c - \hat{\zb u}_\meas }_2^2 + \lambda \norm{ \zb c}_2^2 ,
    \end{align*}
    where $\lambda\in\IR_+$ controls the amount of regularization that is applied. While regularization reduces the noise amplification this comes at the cost of introducing a regularization dependent bias into the solution. In MPI one most often observes a blurring of the reconstructed images and a decreasing spatial resolution. In practice, it is possible to image a wide range of concentrations by adapting the regularization parameter $\lambda$ to the SNR of the measurement data in order to find the best compromise between noise reduction and blurring.
    
    \item A further way to handle noise amplification in the solution is to reduce the condition number of the system matrix by removing the set of rows (frequency components) with low SNR from the linear system. This can be described by the projections 
    \begin{align*}
        \signatur{&\zb P_{\Theta}}{\IC^K}{\IC^{K_{\Theta}}},\\
        \signatur{&\tilde{\zb P}_{\Theta}}{\IC^{K\times N}}{\IC^{K_{\Theta}\times N}}
    \end{align*}
    from $K$ to $K_{\Theta}$ selected frequency components for the measurement vector respectively the system matrix. Removing frequency components of the system matrix with an $\text{SNR}<\Theta$ reduces noise in the solution as the SNR threshold $\Theta$ increases. At the same time, the spatial resolution degrades, since frequency components of higher mixing order are excluded. This is because the relation between mixing degree and SNR is inversely proportional \cite{Rahmer2012TMI,szwargulski2017influence}. To that end, an increase of the SNR threshold has similar effects on the reconstructed images as an increase of the regularization parameter $\lambda$. In MPI this technique works especially well, since the bulk of the background signal can be removed by frequency selection~\cite{them2016sensitivity}. The SNR threshold method requires a measurement of the SNR, which is usually acquired during the measurement of the system matrix. For details and a proper definition of the SNR we refer to \cite{FrankeHybridMRMPI2016}.
    
    \item A third form of regularization is to restrict the number of iterations $\iota$ for iterative solvers, which also speeds up the reconstruction process. This is also important for real-time imaging with MPI as it was done by \cite{knopp2016online, vogel2017low}. The equivalence of regularization and truncated iterations was shown previously~\cite{fleming1990}. While the accuracy of the solution increases with the number of iterations in the beginning, the solution is fitted to match the noise if the number of iterations becomes too large. The optimal iteration number depends on the specific problem and solver. 
\end{enumerate}
In MPI it is quite common to combine the three aforementioned regularization methods for image reconstruction by
\begin{align}\label{eq:Reco}
    \zb R_\cP(\zb S,\hat{\zb u}) 
    = \textup{solve}_\iota\!\left(
    \underset{\zbs c\in \IR^N_+}{\text{argmin}}
    \norm{\tilde{\zb P}_{\Theta}(\zb S) \zb c - \zb P_{\Theta}\!\left(\hat{\zb u}\right)}_2^2 
    + \lambda \norm{ \zb c}_2^2\right)
\end{align}
with a regularization parameter set $\cP = (\lambda,\Theta,\iota)$. Here, $\textup{solve}_\iota$ stands for any iterative solver restricted to $\iota$ iterations.

\section{Problem Statement} \label{Sec:ProblemStatement}

\begin{figure}[t!]
    \centering
    \includegraphics[width=0.7 \textwidth]{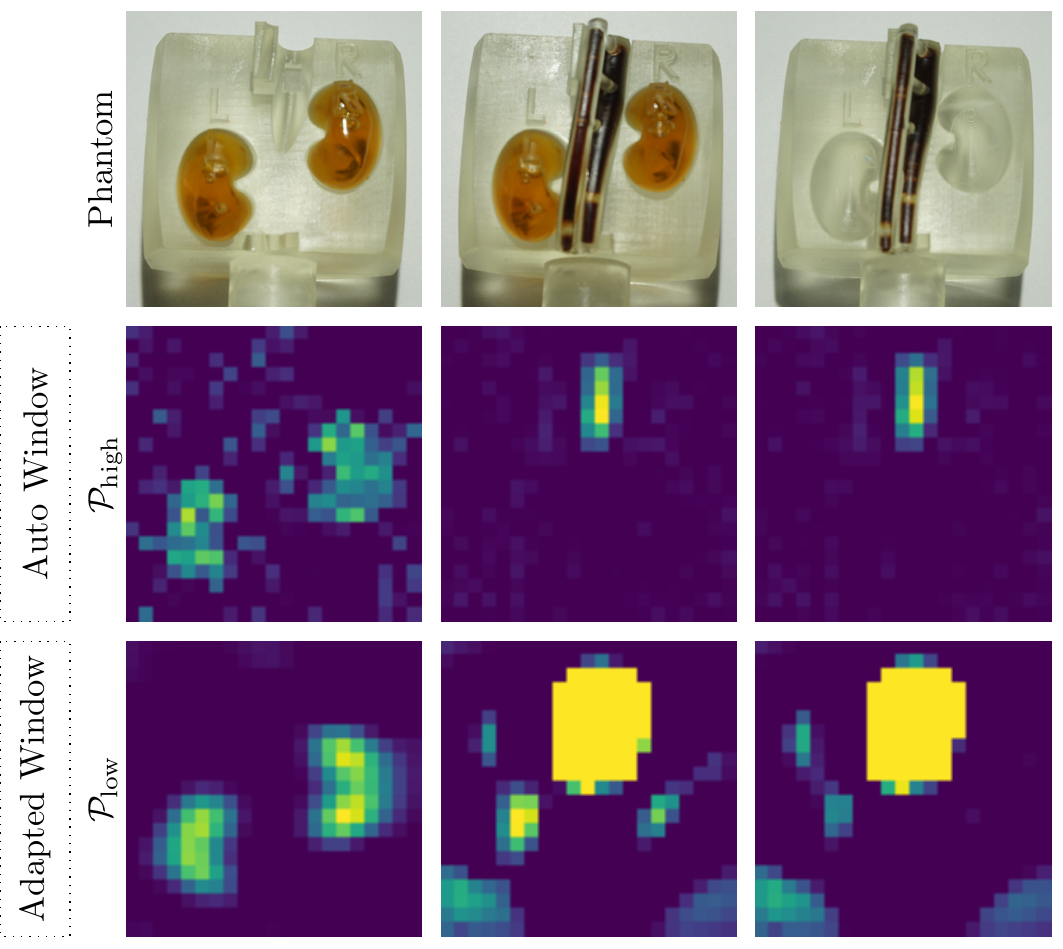}
    \caption{Graphical illustration of the dynamic range limitation in MPI. Shown is a central slice of experimental phantom data measured with a pre-clinical MPI scanner and a 3D Lissajous sequence. Details of experimental parameters are outlined in section \ref{Section:ExpSetup}. In the left column the data from the phantom containing only the kidneys with a concentration of \SI{1.5}{\milli\mol\of{Fe}\per\liter} is shown, while in the central column additional vessels with a concentration of \SI{100}{\milli\mol\of{Fe}\per\liter} are placed between the kidneys. For comparison, the data of the phantom containing only the vessels is shown in the right column. In the first row, the phantoms are depicted, while the second row shows reconstruction results using a reconstruction parameter set adapted to the high concentration of the vessels and windowing parameters adapted to the maximum and minimum value in the data. The third row shows the reconstruction results using the reconstruction and windowing parameters adapted to the concentration of the kidneys.}
    \label{fig:problem}
\end{figure}

The purpose of image reconstruction is to decompose the induced signal into the spatial contribution of particles in the field-of-view (FoV). Ideally, the reconstruction result in one position $\zb r_1$ would be independent of the concentration in a different position $\zb r_2$. In practice, the noise in the measurements and the consequent need for regularization leads to the situation that the reconstruction result in a certain pixel depends on the value in the neighboring pixels. For instance a blurring of the signal can shadow pixels with a smaller value around a pixel with a high value.

To illustrate that this is a real problem in MPI we consider a realistic rat phantom containing kidneys and vessels shown in Fig.~\ref{fig:problem}. The phantom was measured with a 3D Lissajous sequence and scan parameters outlined in section \ref{Section:ExpSetup}. First, the kidneys with an iron concentration of \SI{1.5}{\milli\mol\of{Fe}\per\liter} are measured and reconstructed with appropriate reconstruction parameters, which is shown Fig.~\ref{fig:problem} in the last row on the left. The reconstructed kidneys are clearly visible and thus count as detectable. 
Next, the vessels are placed between the kidneys. Concerning a bolus experiment, they have a concentration of \SI{100}{\milli\mol\of{Fe}\per\liter}, which is a factor of $2^6$ larger than the concentration in the kidneys. In a first step, we reconstruct the data using a parameter set adapted to the high concentration of the vessels and windowing parameters focussing on the entire value range, which is shown in the second row. Here, one can only see the vessels, which is obviously the case since the chosen colorbar has a linear color range and thus changes by a factor of $2^6$ can hardly be detected. Note, that only a part of the vessels is visible since we focus on the image plane where the kidneys are located. The reconstruction parameter are capable of visualizing the kidneys although the reconstruction result is very noisy, which is shown in the second row on the left. Therefore, we provide a second image where the reconstruction parameters and color mapping is chosen to focus on the concentration in the kidneys. This is depicted in the central image in the last row. It is hard to decide if there are just artifacts around the vessels or the kidneys are visible. Therefore, we reconstructed the phantom containing only the vessels with the same reconstruction parameter. As it is shown on the right in the last row, there is no major difference between the reconstructions with and without kidneys. Although we are able to get a good reconstruction result of the kidneys only we are unable to reconstruct them if another object with a much higher concentration is placed in the FoV.

\section{Image Reconstruction for Increased Dynamic Range} \label{Sec:ContrastEnhanced}

Regularization alone can not solve the issue that an object with a high concentration $\kappa_\high$ shadows objects with low concentrations $\kappa_\low$. Therefore, we propose a novel two-step reconstruction algorithm to separate the concentration $\zb c$ into $\zb c_\high$ and $\zb c_\low$, such that
\begin{align*}
    \zb c = \zb c_\high + \zb c_\low.
\end{align*}
Regions with high concentrations should be included in $\zb c_\high$ such that the objects with a concentration of $\kappa_\low$ are visible in $\zb c_\low$. Our approach is based on the linearity of the imaging equation
\begin{align}\label{eq:LinearityImagEq}
    \hat{\zb u}_\meas 
    = \zb S \zb {c} 
    = \zb S (\zb {c}_\high+\zb {c}_\low)
    = \hat{\zb u}_\high+\hat{\zb u}_\low,
\end{align}
which relates $\zb c_\high$ and $\zb c_\low$ with the signals $\hat{\zb u}_\high$ and $\hat{\zb u}_\low$ in raw data space. First, we estimate an image $\zb c_\high$ of the higher concentration $\kappa_\high$, project it to raw data space to obtain an estimate for $\hat{\zb u}_\high$ and calculate $\hat{\zb u}_\low = \hat{\zb u}_\meas - \hat{\zb u}_\high$. In the second step, we reconstruct an image of the lower concentration. The proposed two-step algorithm is outlined in Alg.~\ref{Alg:ContrastEnhanceReco}  
and is described in more detail next:
\begin{enumerate}[label=\arabic*.)]
    \item In the first step, a reconstruction with regularization parameter set $\cP_\high = \left(\lambda_\high, \Theta_\high,\iota_\high\right)$ is performed by solving equation~\eqref{eq:Reco}
    \begin{align*}
        \zb c_\pre = \zb R_{\cP_\high}\!\left(\zb S, \hat{\zb u}_\meas\right).
    \end{align*}

    It results in a relative sharp image $\zb c_\pre$ with an high image noise such that the regions with lower concentrations are shadowed. 
    \item The purpose of the second step is to isolate the area where highly concentrated particles are located. To this end, we perform thresholding in image space with a predefined threshold $\Gamma\in[0,1]$. For that, we define the vector-valued function $\signatur{\bPhi_\Gamma}{ \IR^N }{ \IR^N }$ with the entries $(\bPhi_\Gamma(\zb c))_n$, $n \in I_N$, given by

    \begin{align*}
        (\bPhi_\Gamma(\zb c))_n = \begin{cases} c_n, & \text{if}\; \abs{c_n} \geq  \Gamma \norm{\zb c}_\infty\\
                                       0, & \text{else} ,
                                    \end{cases}
    \end{align*}
    and apply $ \zb c_\thresh = \bPhi_\Gamma(\zb c_\pre)$. The thresholded image $\zb c_\thresh$ is an approximation to the concentration distribution $\zb c_\high$.
    \item In the third step we use the thresholded image and project it to raw data space 
    \begin{align*}
        \hat{\zb u}_\proj = \zb S \zb c_\thresh\in\IC^{K}
    \end{align*} 
    to obtain an approximation to $\hat{\zb u}_\high$.  
    \item In the fourth step we subtract these data from the actual measurement data 
    \begin{align*}
        \hat{\zb u}_\corr = \hat{\zb u}_\meas - \hat{\zb u}_\proj \in \IC^{K}.
    \end{align*}
    Ideally, this will remove the entire part $\hat{\zb u}_\high$ from the measurement data $\hat{\zb u}_\meas$. In practice, one can expect some remaining parts since the reconstruction performed in step one had a non-zero residual. Furthermore, the choice of the thresholding parameter in step two influences the projected data and thus also the residual. 
    \item After subtraction, we reconstruct the corrected signal $\hat{\zb u}_\corr$ via
    \begin{align*}
        \zb c_\post = \zb R_{\cP_\low}\!\left(\zb S, \hat{\zb u}_\corr\right).
    \end{align*}
    In this case we use a regularization parameter set $\cP_\low = (\lambda_\low,\Theta_\low,\iota_\low)$ such that the low concentration part is adequately reconstructed. Therefore, we use higher parameter $\lambda_\low > \lambda_\high$ and $\Theta_\low > \Theta_\high$.
    \item Finally, we add the thresholded high concentration part to the reconstructed low concentration part and return it as the final reconstruction result 
    \begin{align*}
        \zb c_\final = \zb c_\post + \zb c_\thresh.
    \end{align*}
\end{enumerate}
Note that for an efficient implementation, the system matrix in the third step should be projected to $\IC^{K_{\Theta_\low}\times N}$. Since only the projected data is used in the second reconstruction, we can directly use the projected system matrix during the matrix-vector multiplication to reduce computation time and memory consumption.

\begin{algorithm}[t]
    \caption{Proposed two-step reconstruction algorithm for increased dynamic range MPI reconstruction.}
    \label{Alg:ContrastEnhanceReco}
    \SetCommentSty{text}
    \SetCommentSty{footnotesize}
    \KwIn{Acquired data $\hat{\zb u}_\meas$, system matrix $\zb S$, threshold $\Gamma$, regularization parameter sets $\cP_\high = (\lambda_\high,\Theta_\high,\iota_\high)$, $\cP_\low = (\lambda_\low,\Theta_\low,\iota_\low)$}
    \KwOut{Reconstructed particle concentration $\zb c_\final$}
    \Begin{
    \BlankLine
        1. $\zb c_\pre \leftarrow \zb R_{\cP_\high}\!\left(\zb S, \hat{\zb u}_\meas\right)$ \\
        
    	2. $\zb c_\thresh \leftarrow \bPhi_\Gamma(\zb c_\pre) $ \\
    	
    	3. $\hat{\zb u}_\proj \leftarrow \zb S \zb c_\thresh$ \\
    	
    	4. $ \hat{\zb u}_\corr \leftarrow \hat{\zb u}_\meas - \hat{\zb u}_\proj$ \\
    	
        5. $\zb c_\text{post} \leftarrow \zb R_{\cP_\low}\!\left(\zb S, \hat{\zb u}_\corr\right) $ \\
    	
    	6. \Return $\zb c_\post + \zb c_\thresh$ 	
    }
\end{algorithm}

\section{Experimental Setup}\label{Section:ExpSetup}

To investigate the limitation of the dynamic range we performed\textit{ in-vitro} experiments with a pre-clinical MPI scanner (Bruker, Ettlingen, Germany). We applied a 3D imaging sequence with a Lissajous-type sampling trajectory, a gradient strength of $(-0.75,-0.75,1.5)$\,\si{\tesla \per \meter \per \mup} and drive-field amplitudes of \SI{12}{\milli \tesla \per \mup} in each direction. The resulting FoV was of size \SI{32 x 32 x 16}{\milli \meter} and the repetition time of one drive-field cycle was about \SI{21.5}{\milli \second}. The data were averaged over $1000$ drive-field cycles in each case. In all experiments we used a tracer of multi-core superparamagnetic iron-oxide nanoparticles in a dextran matrix (perimag, micromod Partikeltechnologie GmbH, Rostock, Germany, saturation magnetization $\SI{90}{\A\m\squared\per\kg}$ iron for $H > \SI{1}{\tesla\per\mup}$) in different concentrations.

Since the MPI scanner has a non-negligible background signal we performed a background scan with an empty scanner bore and subtracted this measurement from all measurements \cite{them2016sensitivity}. We note that this only removes the static part of the background while due to scanner drifts there still can be parts that cannot be removed by simple subtraction \cite{straub2018joint}.

The system matrix was measured using a robot-based calibration procedure and a delta sample of size \SI{2 x 2 x 1}{\milli \meter} filled with diluted perimag, which has an iron concentration of \SI{300}{\milli \mol \of {Fe} \per \liter}. The robot was moved along \num{21 x 21 x 24} positions located on a regular grid covering a volume of \SI{42 x 42 x 24}{\milli \meter}. The SNR of the system matrix was calculated as it is shown in \cite{FrankeHybridMRMPI2016} with a background correction described in \cite{Knopp2019BG}.

\subsection{Phantoms}
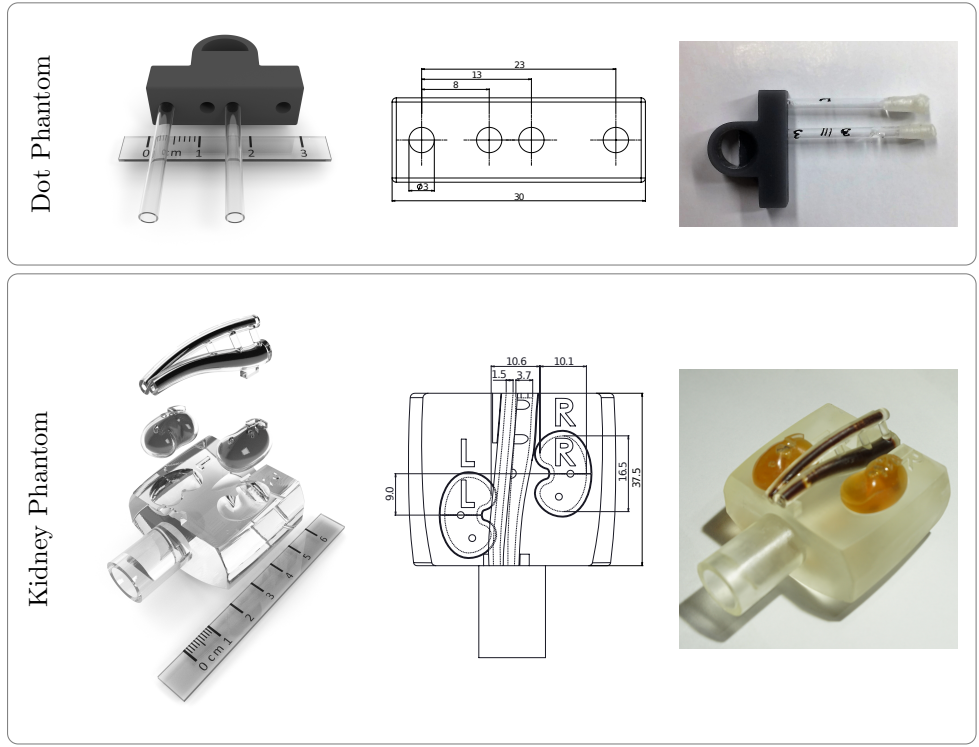
\begin{figure}
    \centering
    \input{tikz/tikz_Phantoms}
    \caption{Phantoms of both experiments. In the first row, the dot phantom is depicted, which consists of a holder and capillaries. Four holes in the holder can be used to place glass capillaries filled with \SI{20}{\micro \liter} tracer material at defined positions. The edge-to-edge distances from the left hole to the remaining three holes are \SIlist{5;10;20}{\milli \meter}. In the second row, the anatomic rat kidney phantom is shown. It consists of a holder where two kidneys and a part of two vessels can be placed in. Each kidney can be filled with \SI{781}{\micro\liter} and the vessels with \SI{306}{\micro\liter} tracer material.}
    \label{fig:phantom}
\end{figure}

\begin{table}
    \begin{center}
    \caption{Concentrations in the capillaries of the phantoms and reconstruction parameters $\cP = (\lambda,\Theta,\iota)$ chosen for each concentration of the dilution series.}
    \label{Tab:Concentrations}
      \begin{tabular}{S S S S S S S}
      \toprule
        {Sample} & {Concentration $\kappa$}  & {Dilution} & \multicolumn{3}{c}{Parameter $\cP$} & {Threshold}\\
        {Number $i$} & {[\si{\milli \mol \of{Fe} \per \liter} ] }& & {$\lambda$} & {$\Theta$} & {$\iota$} & {$\Gamma$}\\\midrule
        1 & 400 & 1 & 0.0005 & 5 & 1 & 1.0\\ 
        2 & 200 & 2 & 0.004 & 5 & 1 & 0.6\\ 
        3 & 100 & 4 & 0.02 & 5 & 1 & 0.3\\ 
        4 & 50 & 8 & 0.06 & 5 & 1 & 0.2\\ 
        5 & 25 & 16 & 0.1 & 5 & 1 & 0.01\\ 
        6 & 12.5 & 32 & 0.3 & 8 & 1 & 0.05\\ 
        7 & 6.25 & 64 & 0.5 & 9 & 1 & 0.03\\ 
        8 & 3.13 & 128 & 0.9 & 10 & 1 & 0.02\\ 
        9 & 1.56 & 256 & 1.4 & 10 & 1 & 0.01\\ 
        10 & 0.78 & 512 & 2.2 & 15 & 1 & 0.005\\ 
        11 & 0.39 & 1024 & 3.2 & 30 & 1 & 0.003\\ 
        12 & 0.2 & 2048 & 4.5 & 40 & 1 & 0.002\\ \bottomrule
      \end{tabular}
    \end{center}
\end{table}

\begin{table}
    \begin{center}
    \caption{Concentrations in the kidneys and the vessel and reconstruction parameters $\cP = (\lambda,\Theta,\iota)$ chosen for each concentration of the dilution series.}
    \label{Tab:ConcentrationsKidney}
      \begin{tabular}{S S S S S S S}
      \toprule
        {Kidney} & {Concentration $\kappa$}  & {Dilution} & \multicolumn{3}{c}{Parameter $\cP$} & {Threshold} \\
        {Number $i$} & {[\si{\milli \mol \of{Fe} \per \liter} ] }& & {$\lambda$}  & {$\Theta$} & {$\iota$} & {$\Gamma$} \\\midrule
        1 & 3.05 & 32 & 0.5 & 15 & 1 & 0.05\\ 
        2 & 1.525 & 64 & 1 & 15 & 1 & 0.025\\ 
        3 & 0.763 & 128 & 2 & 15 & 1 & 0.013\\ 
        4 & 0.381 & 256 & 4 & 24 & 1 & 0.006\\ 
        5 & 0.191 & 512 & 11 & 30 & 1 & 0.003\\ 
        6 & 0.095 & 1024 & 20 & 30 & 1 & 0.002\\ 
        7 & 0.048 & 2048 & 30 & 45 & 1 & 0.001\\ 
        \midrule
        {Vessel} & 100 & 1 & 0.001 & 2 & 3 & {-} \\ \bottomrule
      \end{tabular}
    \end{center}
\end{table}

We used two different phantoms in our experiments. First, a simple concentration phantom was designed using glass capillaries with an inner diameter of \SI{2.4}{\milli \meter} filled with \SI{20}{\micro \liter} tracer material. In the phantom holder, two glass capillaries can be placed with an edge-to-edge distance of \SIlist{5;10;20}{\milli\meter}, which is shown in the first row of Fig.~\ref{fig:phantom}. We prepared a dilution series with $12$ dilutions starting from \SIrange{400}{0.2}{\milli \mol \of {Fe} \per \liter}. In each step, the concentration was halved. Table~\ref{Tab:Concentrations} lists all concentrations used and the associated sample number and reconstruction parameters.
As a second phantom, we used a part of a realistic rat phantom \cite{Exner2019} printed with a Form 3 and clear resin (Formlabs Inc., Somerville, USA). It consists of both kidneys and segments of two vessels as it is shown in the second row of Fig.~\ref{fig:phantom}. While the vessels were filled with \SI{306}{\micro\liter} tracer material with a concentration of \SI{100}{\milli\mol\of{Fe}\per\liter}, each kidney was filled with \SI{781}{\micro\liter} tracer material with a concentration ranging from \SIrange{3.05}{0.048}{\milli\mol\of{Fe}\per\liter}. In each step, the concentration of the kidneys was halved. In Table~\ref{Tab:ConcentrationsKidney}, all concentrations with corresponding reconstruction parameters are listed.

For both phantoms, we first study the idealized dynamic range of the MPI system by measuring samples of different concentration individually. For the dot phantom, the individual samples of the dilution series were moved sequentially into the FoV of the scanner and measured. During reconstruction the SNR threshold $\Theta$ and the regularization parameter $\lambda$ were optimized to provide good SNR and low blurring for each individual measurement by visual inspection. The regularization parameters found are summarized in Table~\ref{Tab:Concentrations}. These parameters were also used as $\cP_\low^i$ for the proposed increased dynamic range algorithm. The parameters for sample $1$ define $\cP_\high$, where the number of iterations was set to $\iota_\high = 50$ to achieve convergence, which leads to low reconstruction bias in the pixels containing the high concentration sample. In the same way, we measured each pair of kidneys and the vessels separately. The reconstruction parameters are listed in Table~\ref{Tab:ConcentrationsKidney}, where the parameters of the kidneys define $\cP_\low^i$ and the parameters of the vessels define $\cP_\high$. For the reconstructions of the kidneys, $\lambda_\low^i$ is scaled with $10^{-1}$ for a better regularization. 

Next, we determined the dynamic range for both phantoms. Using the dot phantom, we measured the highest concentrated sample $1$ together with lower concentrated samples starting from dilution $1$ to $12$. The samples were separated by \SIlist{5;10;20}{\milli \meter} edge-to-edge distance. For each combination of concentration and distance a dedicated measurement was performed. For the kidney phantom, each pair of kidneys was measured together with the vessels.

\subsection{Image Reconstruction}
Each 3D image reconstruction was done by solving Eq.~\eqref{eq:Reco} using an adapted form of the Kaczmarz algorithm \cite{Kaczmarz1937}, to which we refer with \textit{regular reconstruction} in the following. It is implemented in the programming language Julia (version 1.5) \cite{bezanson2017julia} using the package \texttt{MPIReco.jl} (version 0.3.2) \cite{knopp2019mpireco}\footnote[7]{An example of our proposed method is provided at \url{https://github.com/IBIResearch/IncreasedDynamicRange}}. Note that the regularization parameter $\lambda$ is scaled with $\textup{trace}(\tilde{\zb P}_{\Theta}(\zb S)^H \tilde{\zb P}_{\Theta}(\zb S))N^{-1}$ in the implemented reconstruction method \cite{Weizenecker2007PhysMedBio}.

Once the reconstruction parameters were determined for the idealized dynamic range phantom, we performed three reconstructions for the two-dot dynamic range phantom. First, each dilution level was reconstructed using $\cP_\high$ and second using the corresponding set $\cP_\low^i$ for the lower concentration sample. Third, we applied the proposed two-step algorithm, which requires both sets of parameters $\cP_\low^i$ and $\cP_\high$. The threshold parameter was chosen to isolate all data with a signal slightly higher than the concentration of the lower concentrated sample. In order to achieve this, we set $\Gamma_i=(0.8^{i-1})^{-1}$ for the dot phantom and for the realistic kidney phantom we used the thresholds $\Gamma_i=0.05 (2^{i-1})^{-1}$, which are also listed in Table~\ref{Tab:Concentrations} and \ref{Tab:ConcentrationsKidney}. Additionally, reconstructions with varying thresholds $\Gamma_j = 0.5 (2^{j-1})^{-1}$, $j\in\set{1,\dots,7}$ for selected concentrations were performed for the dot phantom.

\subsection{Signal-to-Artifact Ratio}
Apart from visual analysis of the reconstructed images, we quantified the signal-to-artifact ratio (SAR) in the final images of the dot phantom. 
The aim is to set the artifact level in relation to the measured particle concentration, since discrimination can only happen if the artifacts are smaller than the particle concentration. For the calculation of the SAR, we define two masks.
First, $\zb M_\signal \in \set{0,1}^{N \times N}$ is a diagonal matrix, which describes the voxels covered by the lower concentrated sample.  
The second mask is the diagonal matrix $\zb M_\artifact \in \set{0,1}^{N\times N}$, which describes the region without any sample. 
Then, the SAR is set to the maximum value of the reconstructed concentration of the sample with lower iron amount by the maximum of the intensities of the voxels without any sample:

\begin{align*}
    \text{SAR}(\zb c) = 
    \frac{\norm{\zb M_\signal \zb c}_\infty}
    {\norm{\zb M_\artifact \zb c}_\infty}.
\end{align*}
 
A ratio of $1$ corresponds to artifacts in the same range as the amplitude of the lower concentrated sample and thus discrimination is impossible.

\section{Results}\label{Sec:Results}

The reconstruction results of the presented phantoms are described next. For both phantoms we distinguish between the idealized dynamic range experiments where only the concentration sample is placed in the scanner and the dynamic range experiments where the high and the low concentrated samples are imaged together.

\setlength{\fboxrule}{2pt}
\setlength{\fboxsep}{0pt}

\subsection{Dot Phantom}

\begin{sidewaysfigure*}
\vspace{13cm}
    \centering
    \includegraphics[width=0.9\textwidth]{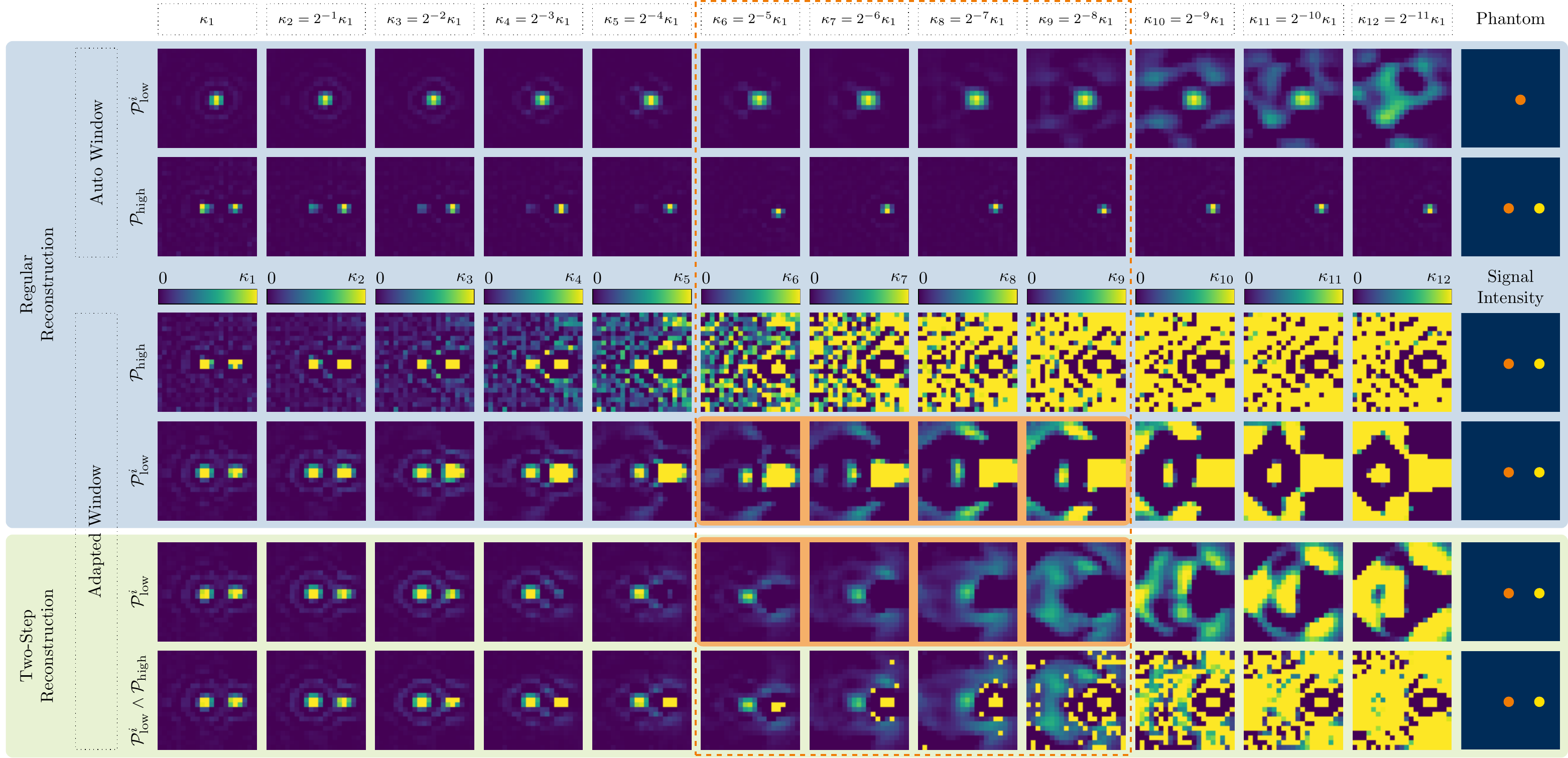}
    \caption{\small Reconstruction results of the dilution series with a single sample in the first row and two samples with a distance of \SI{10}{\milli\meter} in the other rows. In each experiment, a single capillary filled with \SI{20}{\micro \liter} tracer material of distinct concentrations given in Table~\ref{Tab:Concentrations} is located in the FoV. The concentrations decrease by a factor of $2$ from left to right. Each image shows the maximum intensity projection over the $7$th to $9$th $xy$-plane of the reconstructed 3D volume. In the experiment with two samples, the additional sample on the right has the concentration $\kappa_1$ in each experiment. The first four rows show the reconstruction results with the regular reconstruction. In the single-sample experiment, we used the parameter set $\cP_\low^i$. The second and third row show the two-sample experiment reconstructed with $\cP_\high$ using an auto respectively an adapted window clipped at the concentration of the lower concentrated sample. The reconstruction results with $\cP_\low^i$ using the adapted window is shown in the fourth row. In the last two rows, the results of the proposed two step reconstruction with the adapted window are shown. First, $\zb c_\post$ is shown in the fifth row using $\cP_\low^i$. In the last row, the sum $\zb c_\thresh + \zb c_\post$ using both parameter sets is shown. The orange rectangles mark the reconstruction results, which are compared to the results for other distances of the samples in Fig.~\ref{Fig:TowDotsPhantomDist}.}
    \label{Fig:TowDotsPhantom}
\end{sidewaysfigure*}

\subsubsection*{Idealized Dynamic Range:}

First, we summarize the results of the experiments to determine the idealized dynamic range, in particular the lower bound, with a single sample inside the measuring field. The first row of Fig.~\ref{Fig:TowDotsPhantom} shows the reconstruction results for each of the single-sample phantoms. It was possible to image the sample for the concentrations $\kappa_1$ -- $\kappa_{11}$ while the sample cannot be recognized for $\kappa_{12}$. The iron content of the $11$th sample was \SI{440}{\nano \gram \of {Fe}}, which is in similar range as reported in \cite{Graeser2017SR} for the same imager ($\approx$ \SI{160}{\nano \gram \of {Fe}} but with a different tracer material generating about $3$ times more signal). The quality of the reconstructed particle concentration is similar for concentrations $\kappa_1$ -- $\kappa_{8}$. Starting from $\kappa_6$ one can observe a blurring of the dot, which is due to the increasing regularization. In addition, artifacts that already appeared in former images, become prominent for concentrations $\kappa_{10}$ and $\kappa_{11}$. This is reflected by the SAR summarized in Fig.~\ref{Fig:artifactLevel}. Overall the SAR decreases for $\kappa_1$ -- $\kappa_{11}$ until it drops below $1$ for $\kappa_{12}$.

\subsubsection*{Dynamic Range:}
    
Next, we investigate the results for the dilution series with two samples for the determination of the dynamic range. For an edge-to-edge distance of \SI{10}{\milli\meter}, the results are shown in Fig.~\ref{Fig:TowDotsPhantom}. In the second row, where the parameter set $\cP_\high$ is used, the sample with the highest concentration is visible in all reconstructed images and appears with only slight variations in the intensity of its pixel values. The sample with varying concentrations is only visible for concentrations $\kappa_1$ -- $\kappa_4$. For lower concentrations one cannot discriminate its signal from background noise anymore.
    
In order to focus on the low concentration sample we adapted the windowing in such a way that the maximum window level is derived from the single-sample dilution series. The results are shown in the third row. We note that the high signal values are clipped due to the adapted windowing. In the first five images, both samples can be clearly detected. For the first concentrations $\kappa_1$ -- $\kappa_3$ only slight image artifacts are visible. Starting from $\kappa_4$ one can identify the background noise, which is in the same range as the low concentration sample for $\kappa_6$. 
    
To reduce the background noise one can use the regularization parameter sets $\cP_\low^i$. The results are shown in the fourth row of Fig.~\ref{Fig:TowDotsPhantom} with the same windowing as before. The background noise is strongly reduced, but instead ring-like background artifacts occur. In the first $6$ reconstruction results both samples can be seen. The larger the regularization parameters, the more blurred is the highly concentrated sample. From concentration $\kappa_7$ on, the lower concentrated sample is widened vertically and moves to the left, which will be more visible later when comparing the three distances. Additionally, the ring-like artifact around the lower concentrated sample gets more prominent.

Finally, the last two rows of Fig.~\ref{Fig:TowDotsPhantom} show the results of our proposed algorithm for increased dynamic range. As for the third and the fourth row, the color mapping was adapted to fit the lower concentration. We show both, the reconstruction result $\zb c_\post$, where the highly concentration part was subtracted, and the final image $\zb c_\final$, where the latter has been added.

As one can see in the images, the algorithm is capable of reconstructing the low concentration sample until concentration $\kappa_8$. Since the deviation between both concentrations is too small, the higher concentrated sample is also visible for $\kappa_1$ -- $\kappa_3$. Below, one can see a blue area in those parts, where the high concentration sample was located. Only in some cases a small signal is present. This shows that the removal of the high concentration signal in raw data space does work. Around this blue area one can still identify a ring-like artifact, which gets more prominent for lower concentrations and can be attributed to the high concentration signal.
For concentration $\kappa_8$ the lower concentrated sample starts to vanish into the artifact but they are still distinguishable. At $\kappa_9$, the reconstruction result gets ambiguous. It is also visible that the dot in the $7$th and $8$th image is not at the same position as in the regular reconstruction result in the row above. For all lower concentrations $\kappa_{10}$ -- $\kappa_{12}$ one cannot identify the sample with lower concentration anymore. In the final reconstruction results in the last row, a noise-like artifact emerges from concentration $\kappa_7$ on. This can be traced back to the lower threshold $\Gamma$, which also selects voxels with artifacts for $\zb c_\thresh$.

\begin{figure}
    \centering
    \includegraphics[width=0.99\textwidth,trim=0.8cm 0cm 0cm 0cm, clip]{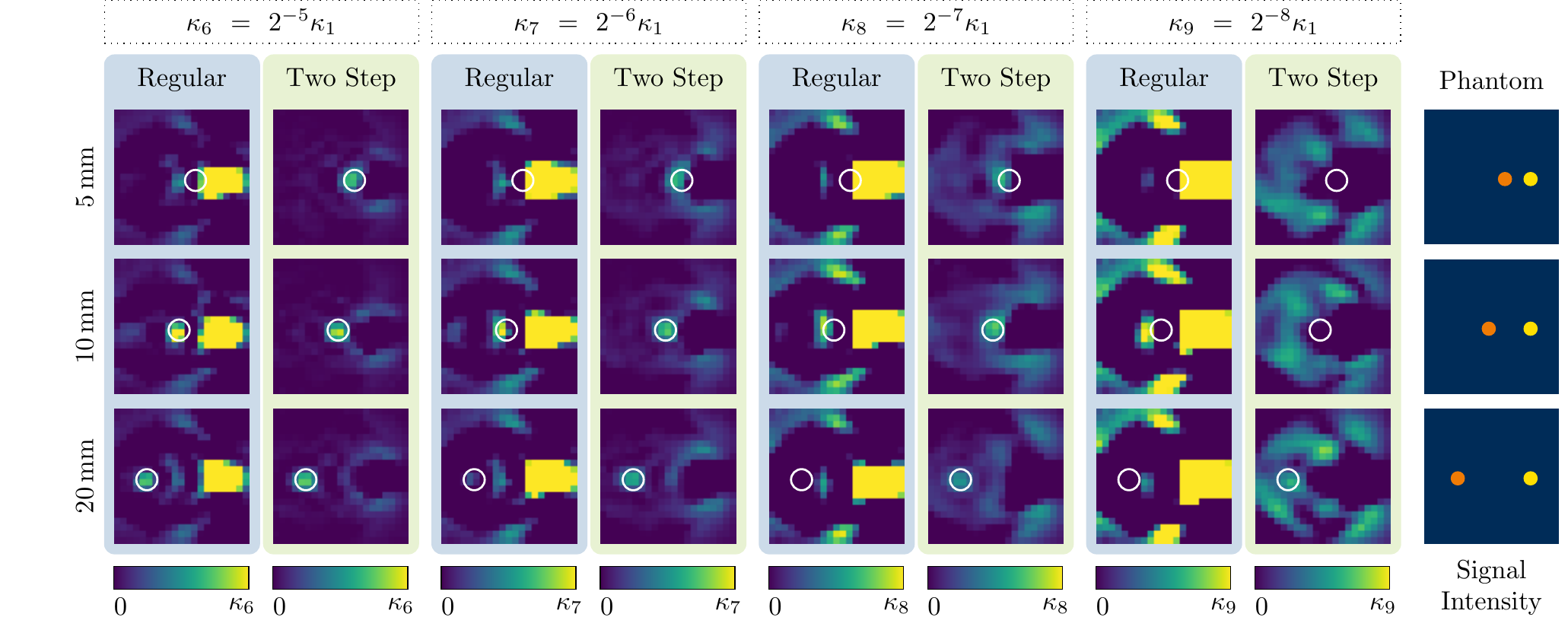}
    \caption{Reconstruction results of the two-sample phantom for different sample distances \SIlist{5; 10; 20}{\milli\meter}. Shown are the reconstruction results for the chosen concentrations $\kappa_6$ -- $\kappa_9$. For each concentration, one column shows the reconstruction results using the regular reconstruction while the other column shows the results $\zb c_\post$ using the two-step reconstruction. All reconstructions use the parameter set $\cP_\low^i$ and afterwards the adapted window is clipped at the concentration of the lower concentrated sample. The position of the sample is marked by a white circle in each image.}
    \label{Fig:TowDotsPhantomDist}
\end{figure}

For the concentrations $\kappa_6$ -- $\kappa_9$, the reconstruction results for the three distances \SIlist{5;10;20}{\milli\meter} using the parameter sets $\cP_\low^i$ and the regular reconstruction method are shown in Fig.~\ref{Fig:TowDotsPhantomDist} in the blue columns. The green columns show the reconstruction result $\zb c_\post$ using the two-step reconstruction. The white circle indicates in each image the correct position of the sample with the lower iron amount. The \SIlist{10;20}{\milli\meter} samples are clearly visible for $\kappa_6$ for both reconstruction methods, while the \SI{5}{\milli\meter} sample is only visible in the two-step reconstruction result. The enlarged higher concentrated sample reaches into the position of the \SI{5}{\milli\meter} sample. Additionally, an artifact occurs in the regular reconstruction result, which appears also in the result of the \SI{20}{\milli\meter} sample. For lower concentrations this artifact gets more prominent such that the sample vanishes. From $\kappa_8$ on, there is no sample detectable at the correct position for the regular reconstruction method. In the two-step reconstruction, the \SIlist{10;20}{\milli\meter} sample is clearly visible until $\kappa_8$. Only the \SI{5}{\milli\meter} sample changes slightly its position.

\begin{figure}
    \centering
    \includegraphics[width=0.99\textwidth,trim=0.0cm 0cm 0cm 0cm, clip]{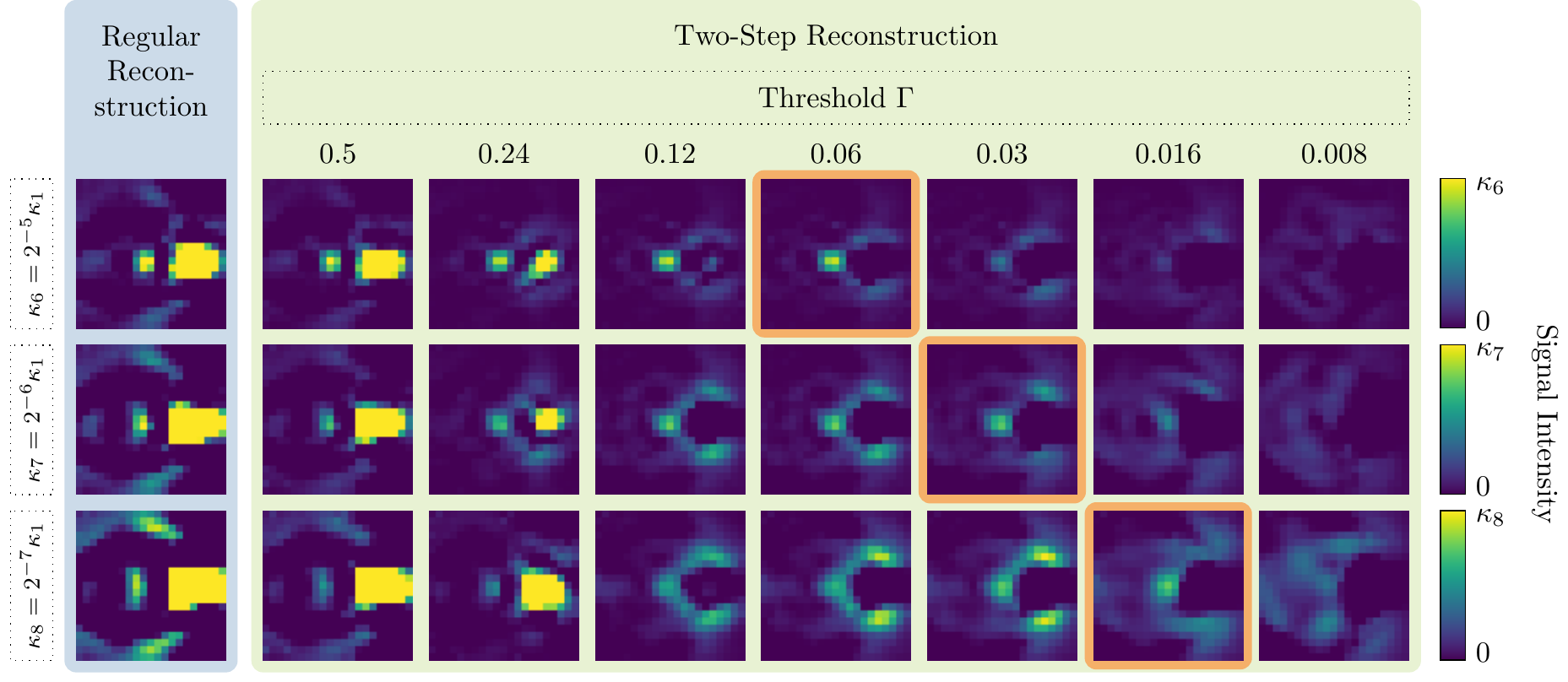}
    \caption{Impact of the threshold $\Gamma$ on the results of the two-step reconstruction method. Shown are the reconstruction results $\zb c_\post$ of the two-sample phantom with a distance of \SI{10}{\milli\meter} of the samples for different thresholds and selected concentrations $\kappa_6$ -- $\kappa_8$. The window is adapted to the concentration of the sample with a lower iron amount. For comparison, the results of the regular reconstruction method with parameter set $\cP_\low^i$ are shown on the left. The thresholds used in Fig.~\ref{Fig:TowDotsPhantom} are marked with an orange square.}
    \label{Fig:TowDotsPhantomThresh}
\end{figure}

In Fig.~\ref{Fig:TowDotsPhantomThresh}, reconstruction results for $\kappa_6$ -- $\kappa_8$ for varying thresholds $\Gamma$ are shown. Starting with $\Gamma = 0.5$ in the second column, the results of the two-step reconstruction method look similar to the results of the regular reconstruction in the first column. While $\Gamma$ drops, the signal of the sample with higher iron amount vanishes. The orange rectangles mark the best choice of $\Gamma$ for each concentration. Approximately these values are also used in Fig.~\ref{Fig:TowDotsPhantom}. If $\Gamma$ is chosen lower than the marked ones, the signal of the lower concentrated sample is included in $\zb c_\thresh$ and therefore not reconstructed in the second step of our proposed method.

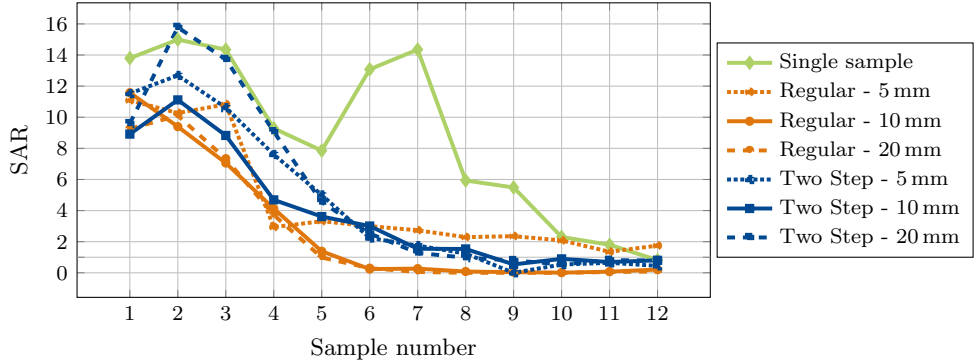
\begin{figure}
    \centering
    \input{tikz/tikz_SAR}
    \caption{Signal-to-artifact ratio of the reconstructed images using the regular and the two-step reconstruction. In the case of the regular reconstruction, the parameter set $\cP_\low^i$ was used. For the two-step reconstruction, the SAR is calculated for $\zb c_\post$. The SAR is determined to be the ratio between the maximum intensity at the spatial location of the sample with lower amount of particles and the maximum intensity in regions without particles. For comparison, the SAR of the reconstructed images for the single sample experiment using the regular reconstruction with $\cP_\low^i$ is provided.}
    \label{Fig:artifactLevel}
\end{figure}

The SARs are summarized in Fig.~\ref{Fig:artifactLevel} comparing the regular reconstruction using $\cP_\low^i$ and $\zb c_\post$ of the two-step reconstruction. Additionally, the SAR of the reconstruction results of the single sample is provided. The SAR indicates if the signal is distinguishable from the artifacts ($> 1$). The behavior of the SARs fits the visual inspection of the reconstructed images. Due to the enlarged high concentrated sample, there is a significant signal at the position of the \SI{5}{\milli\meter} sample in the regular reconstruction result. Thus, the SAR is higher compared to the other reconstruction results for lower concentrations. For the relevant concentrations from $\kappa_5$ -- $\kappa_8$ the SAR of the two-step reconstruction results are significantly higher than the results of the regular reconstruction for the \SIlist{10;20}{\milli\meter} sample. While the SAR of the regular reconstruction results is lower than $1$ from $\kappa_6$ on, the SAR of the two-step reconstruction results is greater than $1$ until $\kappa_8$. An SAR greater than $1$ indicates that the sample can be distinguished from the artifacts, which agrees with the qualitative findings of the reconstruction results shown in Fig.~\ref{Fig:TowDotsPhantom}.

\subsection{Kidney Phantom}

Next, we investigate the performance of the two-step reconstruction algorithm compared to the regular reconstruction algorithm on the kidney phantom. An extended analysis can be found in  \ref{appendixA}. 

\begin{figure}
    \centering
    \input{tikz/tikz_RecoVessel}
    \caption{Reconstruction results of the kidney phantom containing only the vessels. On the left, the phantom with its coordinate system is shown. The vessels are filled with \SI{306}{\micro\liter} tracer with a concentration of \SI{100}{\milli\mol\of{Fe}\per\liter}. On the right, the reconstruction results using the regular reconstruction algorithm are depicted. The reconstruction result on the right shows the $10$th $xy$-plane, where the kidneys are located. The left reconstruction result shows a maximum intensity projection of the $10$th and $11$th $xz$-plane to visualize both vessels. The dotted lines show the intersection of both reconstructed images.}
    \label{fig:recoVessel}
\end{figure}
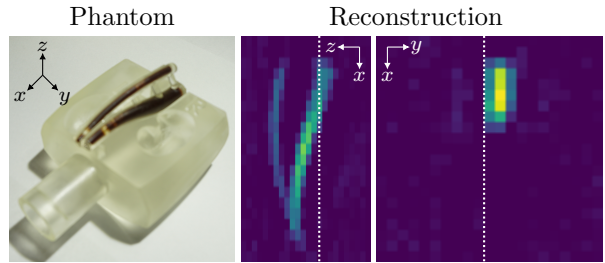

\subsubsection*{Vessels:} Fig.~\ref{fig:recoVessel} shows a reconstruction of the vessels without kidneys. We used the regularization parameter set $\cP_\high$ so that we were able to separate both vessels as shown in the central image. On the right side, the $10$th $xy$-plane is shown, which is the plane where the kidneys are located. Since our focus is on the kidneys, the vessels are represented by the longish dot in the following reconstructed images.

\begin{figure*}
    \centering
    \includegraphics[width=0.99\textwidth]{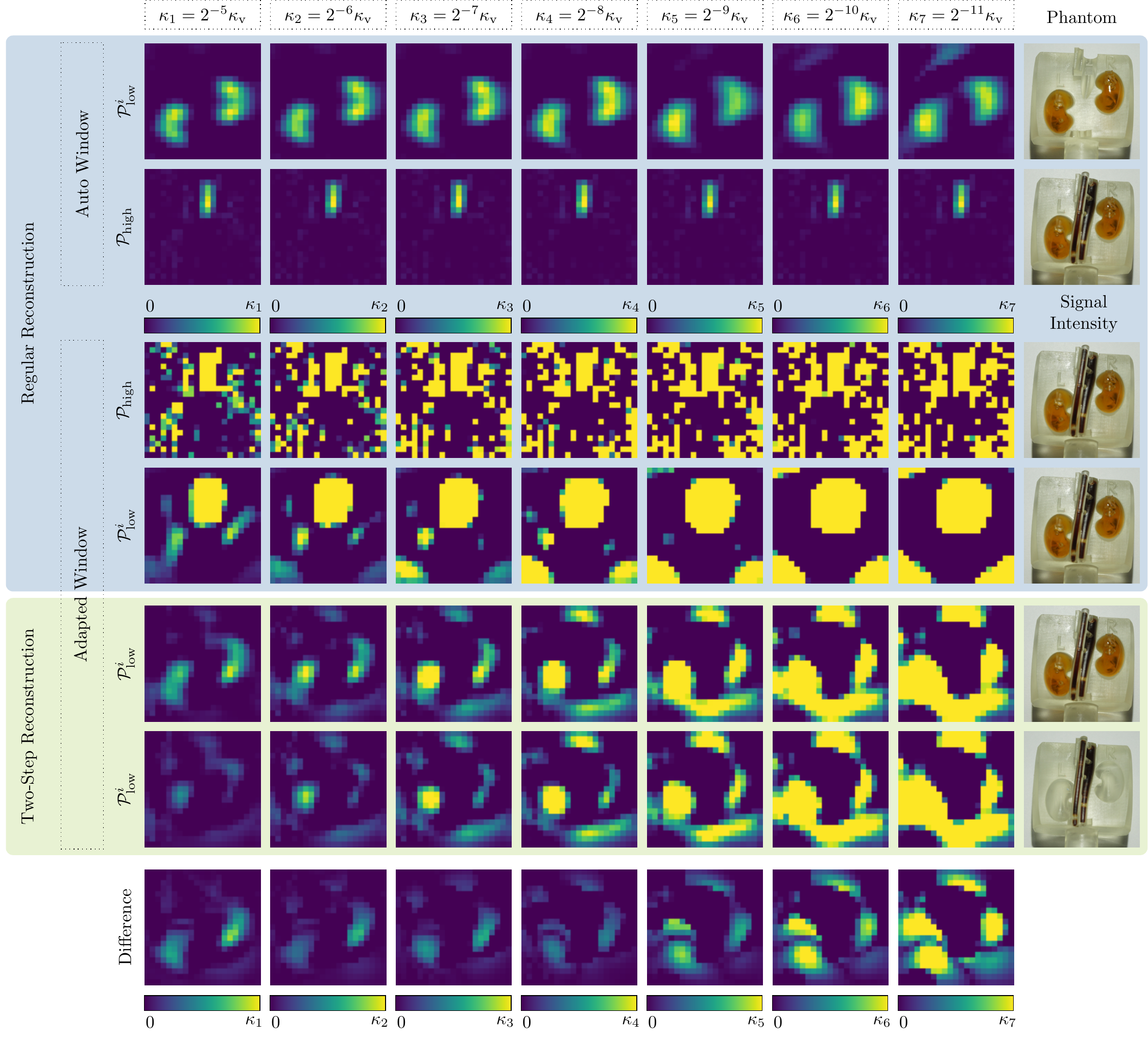}
    \caption{Reconstruction results of the kidney phantom. In each experiment, the vessel is filled with \SI{306}{\micro \liter} tracer material with a concentration of $\kappa_v = \SI{100}{\milli\mol\per\liter}$ while each kidney is filled with \SI{781}{\micro \liter} tracer material of distinct concentrations given in Table~\ref{Tab:ConcentrationsKidney}. The concentrations decrease by a factor of $2$ from left to right. Each image shows the $10$th $xy$-plane of the reconstructed 3D volume. The first four rows show the results for the regular reconstruction. In the first experiment, where only the kidneys were placed in the scanner, we used the parameter set $\cP_\low^i$. The second and third row show the experiment with kidneys and vessel, reconstructed with $\cP_\high$ using an auto respectively an adapted window clipped at the concentration of the kidneys. The reconstruction results with $\cP_\low^i$ using the adapted window is shown in the fourth row. In the next two rows, the results of the proposed two-step reconstruction with the adapted window are shown. First, $\zb c_\post$ is shown in the fifth row using $\cP_\low^i$. For comparison, the results of the two-step reconstructions applied only on the vessel using $\cP_\low^i$ are shown in the sixth row. The absolute values of the differences between the two rows showing the two-step reconstruction results are illustrated in the last row.}
    \label{Fig:KidneyPhantom}
\end{figure*}

\subsubsection*{Kidneys:}
In Fig.~\ref{Fig:KidneyPhantom}, the reconstruction results of the regular and the proposed two-step reconstruction methods applied to the kidney phantom are shown. The first row shows the reconstruction results where just the kidneys are placed in the FoV and the parameter set $\cP_\low^i$ is used. We are able to visualize the kidneys for all concentrations. For $\kappa_1$ to $\kappa_4$ the correct shape of the kidneys could be reconstructed. From $\kappa_5$ to $\kappa_7$ additional artifacts appear but it is still possible to distinguish between artifacts and kidneys.

\subsubsection*{Kidneys and Vessels:}
In the next four rows of Fig.~\ref{Fig:KidneyPhantom}, different reconstruction results for the phantom containing the kidneys and the vessels are shown. First, in the reconstruction results using $\cP_\high$ only the vessels are visible. Adapting the window to the concentration of the kidneys leads to noise in the image such that the kidneys are not clearly visible (third row). Using the parameter set $\cP_\low^i$ of the kidneys for the regular reconstruction yields two longish dots at the position of the kidneys for $\kappa_1$ (fourth row). But the shape of the kidneys is not identifiable. From $\kappa_2$ on, the artifacts in the corners get more distinct while the two dots vanish. In the fifth row, the result $\zb c_\post$ of the proposed two-step reconstruction method is shown where the window is adapted to the concentration of the kidneys. For $\kappa_1$ and $\kappa_2$ both kidneys are clearly visible and also the correct shape is reconstructed for the kidney on the right side. For $\kappa_2$, the artifacts get stronger such that the left kidney already starts to vanish into an artifact. For $\kappa_3$ to $\kappa_7$ the artifacts get too prominent. For comparison, the sixth row shows the two-step reconstruction result $\zb c_\post$ of the phantom containing only the vessels. The absolute value of the difference between the results in the fifth and in the sixth row is shown in the last row. It emphasizes that we are able to visualize the kidneys for $\kappa_1$ and $\kappa_2$ while the artifacts are dominating the other results.

\section{Discussion}\label{Sec:Discussion}

A comparison of the results from the dilution series of a single sample with the dilution series of two samples indicates a limited dynamic range of MPI. In the single-sample dilution series it was possible to image concentrations from $\kappa_1$ -- $\kappa_{11}$ representing a dilution factor of $2^{10} =1024$. This factor is only a lower bound, since we did not increase the iron concentration beyond \SI{400}{\milli \mol \of{Fe}\per \liter}. We note, however, that higher iron concentrations are not feasible since then particle-particle interactions lead to non-linear relation between the particle concentration and the generated signal \cite{loewa2016concentration}. In turn the imaging equation \eqref{Eq:ImagingEquation} would not be fulfilled anymore and thus the regular image reconstruction approach would not be appropriate.
In the experiments for determining the dynamic range using a two-sample phantom, the regular reconstruction algorithm was capable of resolving a dynamic range of about $2^5=32$ for a distance of \SIlist{10; 20}{\milli\meter}. In the \SI{5}{\milli\meter} case we were able to resolve only a dynamic range of about $2^4=16$ due to the artifacts. Thus, there is a gap of at least one order of magnitude. We note that those results were observed for an averaging factor of $1000$, which was applied to lower the detection limit of the scanner by a factor of $\sqrt{1000}\approx 2^5$. For unaveraged and thus more noisy data the idealized dynamic range decreases until it lies in the same order of magnitude as the dynamic range. The problem described in this paper thus is primarily relevant for data with high SNR. 

Using the proposed reconstruction algorithm it is possible to increase the dynamic range to about $2^7=128$ for a distance of \SIlist{10; 20}{\milli\meter} and to about $2^6=64$ for \SI{5}{\milli\meter} distance. Thus, the two-step reconstruction method is able to increase the dynamic range by a factor of $4$. In addition, it also improves the signal-to-artifact ratio for lower concentration differences. The SAR is higher for the proposed algorithm compared to the regular reconstruction algorithm for almost all considered concentrations.

One interesting finding was that the dynamic range increases with the edge-to-edge distance of the samples. This is supported by the fact that the regularization leads to a blurring or ringing like artifact, which is always a local effect.
The regularization parameters for very low concentrations result in a large support of the artifacts covering the entire FoV. Thus, those concentrations ($\kappa_{10}$ -- $\kappa_{12}$) cannot be imaged even for \SI{20}{\milli \meter} edge-to-edge distance. Note that also the full width at half maximum of the point spread function of the particles affects the dynamic range, which can not be influenced by reconstruction methods. Apart from the increased dynamic range, a further advantage of the algorithm is the flexibility to use different regularization parameters for the high and the low concentration part in an image. This enables us to achieve a high resolution for highly concentrated structures (e.g. vessels) in the presence of a lower concentrated structures.

In order to visualize concentration differences larger than a factor of ten, we had to use two different windowing parameters for the color mapping. One was focused on the high concentration, the other was focused on the low concentration. This shows that large concentration differences do not only require special attention in the reconstruction step but also during visualization. One way to handle this is to develop common experience-based window parameters like it is also done in computed tomography. This induces clipping of higher values, which can be avoided by using a non-linear colormap. In addition, poly-chromatic colormaps emphasize fine concentration differences more than mono-chromatic colormaps \cite{moreland2009diverging}.

In the introduced algorithm, we isolate the pixels that belong to the high concentration in the second step. Here, we introduced a simple threshold that decides independently for each pixel whether it belongs to the high concentration or not. This leads to isolated pixel in the thresholded image since noise is selected if the threshold is small. The area with high concentrated tracer is usually connected and thus the finally reconstructed images could be improved with clustering or image segmentation methods to remove noisy pixels in the thresholded high-concentration image.

Within this work we considered a single-patch 3D acquisition sequence. One way to mitigate the dynamic range problem could be to switch to multi-patch image sequences \cite{Knopp2015PhysMedBiol,gdaniec2017fast}. The idea is to split the measuring field into locations with high and low particle concentration and measure them separately. This approach depends strongly on the spatial distribution of tracer material and there are configurations where the separation into rectangular patches is impossible. Since the signal strongly drops if a sample is moved outside the FoV \cite{weber2015artifact}, such a method may enable resolving a higher dynamic range. One interesting question is if the joint reconstruction approach introduced in \cite{Knopp2015PhysMedBiol} that enable continuous patch transitions is still the best approach. It might be better to apply a separate reconstruction of the individual patches as discussed in \cite{ahlborg2016using}. Instead of using 3D Lissajous sampling patterns in combination with multi-patch sequences one could also switch to 2D or 1D excitations \cite{knopp2008trajectory,werner2017first}. This reduces the space that is covered during one drive-field cycle, which might increase the dynamic range for objects that are not arranged along the excitation line. 

The developed algorithm requires only slight modification of the existing reconstruction process and increases the reconstruction time by roughly a factor of two since it involves two inner reconstructions. The algorithm is also not restricted to the separation of two different concentration levels but it is also possible to extend the concept and reconstruct a series of different concentration levels. One would then successively remove parts from the signal such that in the end only the least concentrated image remains. In the end, one could then sum up all reconstructed concentration levels.

Furthermore, the algorithm is not restricted to the regularized system matrix reconstruction as  defined in Eq.~\eqref{eq:Reco}. First of all, other regularization methods besides Tikhonov regularization like the fused lasso model~\cite{Storath2017} can be applied. Also $x$-space reconstructions can be used instead of system matrix reconstructions. Note that this also requires adaption of the back projection into raw data space in the third step of the two-step reconstruction method.

Within this work we started with a simple test phantom consisting of two capillaries to study the dynamic range in an \textit{in-vitro} environment. This scenario is also relevant for \textit{in-vivo} imaging. Based on estimations of the mean blood density in organs, one can estimate a 70 fold lower particle concentration in the brain compared to the concentration in vessels \cite{Oeff1955}. Our experiment showed, that we are able to reconstruct a dynamic range of this magnitude with the two-step reconstruction method. In mice and rats, the contrast material is injected via the tail vein and is purified during the passage of the vascular tree. Organs located closely to the vena cava or heart, like the liver or lungs are thus hard to image in the first pass. To illustrate this scenario, we provided the second realistic phantom consisting of rat-sized vessels and kidneys. We were able to visualize the kidneys for all concentrations down to \SI{48}{\micro\mol\of{Fe}\per\liter}. But even the lowest relation between the vessels and the kidneys, which was a dilution of $2^5=32$, could not be reconstructed with the regular method since the kidneys were not clearly distinguishable from the artifacts. Our proposed method was capable of reconstructing the kidneys with a dynamic range up to $2^6=64$. Thus, the proposed algorithm can help increasing the dynamic range in \textit{in-vivo} experiments to image organs in the first pass.

\section{Conclusion}

In conclusion, we found that the presence of high particle concentrations within the FoV of an MPI system can significantly reduce the dynamic range the system is able to resolve.  This is an issue in pre-clinical applications, where a highly concentrated tracer bolus in the vascular system in certain situations prevents imaging of nearby organs with lower tracer concentration. In this work we proposed a two-step reconstruction algorithm which can increase the dynamic range in the aforementioned scenario and reduce image artifacts. In experiments we  observed an improvement in the dynamic range by a factor of $4$. Similarly, the artifact-to-noise ratio improved by a factor of up to $21$.

\section*{References}

\bibliographystyle{ieeetr} 
\bibliography{ref}

\appendix
\section{Extended Analysis of the Two-Step Reconstruction Method} \label{appendixA}
\begin{figure*}
    \centering
    \includegraphics[width=0.99\textwidth]{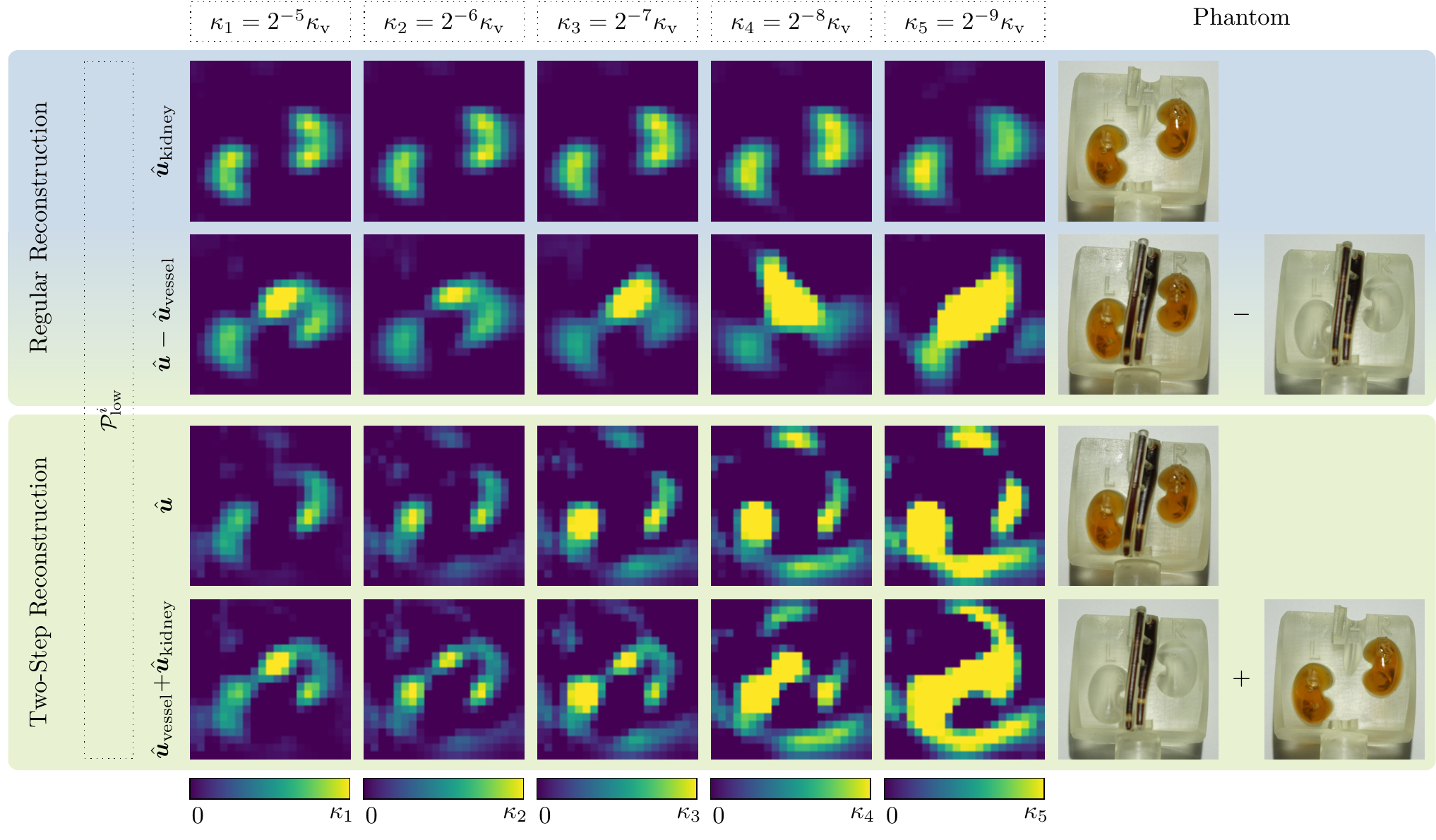}
    \caption{Additional reconstruction results of the kidney phantom. For comparison, the regular reconstruction result of the kidneys and the two-step reconstruction result $\zb c_\post$ of the entire phantom from Fig.~\ref{Fig:KidneyPhantom} are shown in the first and third row. In the second row, the reconstruction result of the entire phantom is shown where the raw data of the vessels is subtracted, which is equivalent to the two-step reconstruction where $\zbh u_\proj = \zbh u_\vessel$. In the last row, the raw data of the separate measurements of vessels and the kidneys are summed up and reconstructed with the two-step method.}
    \label{Fig:KidneyPhantomTesting}
\end{figure*}
In the appendix, two additional reconstructions are shown to illustrate the properties and limitations of the two-step algorithm. The results  were calculated for the kidney phantom are shown in  Fig.~\ref{Fig:KidneyPhantomTesting}.

The first part is to investigate how well the subtraction of the highly concentrated sample can be ideally performed. Since we measured the vessels separately from the kidneys, it is possible to subtract the raw data of the vessels directly from the signal of the composite phantom. This can be seen as an ideal case, where the first step of the two-step procedure does not cause any systematic error, which can occur due to thresholding. The second row in Fig.~\ref{Fig:KidneyPhantomTesting} shows the result of the two-step algorithm if $\zbh u_\proj$ is replaced by $\zbh u_\vessel$, in which case the two-step algorithm effectively becomes the ordinary one-step algorithm. The results are compared to reconstruction results from the kidneys only (first row) and the regular two-step reconstruction algorithm of the entire phantom (third row). One can see that that in this idealized case, the kidneys can be reconstructed even for $\kappa_3$, which is one step further than the reconstruction with $\zbh u_\proj$. Nevertheless, even in this idealized case it is not possible to retrieve the kidneys for the low concentration level $\kappa_5$ and below. An explanation for this could be that very slight changes either in the positioning of the phantom or slight differences in the applied imaging sequence occurred. 

In the second part of the extended study another idealized case is considered. This is the case in which the kidneys and the vessels are measured separately and then digitally added. In this situation, we have the optimal linearity of Eq.~\eqref{eq:LinearityImagEq}. Despite of an artifact in the center, the reconstruction results shown in the last row of Fig.~\ref{Fig:KidneyPhantomTesting} are similar to the results of the composed phantom. While the left kidneys have the same structure for all concentrations the shape of the right kidney is a bit better than in the third row, which can be caused by the artifact.
Although we observe an artifact in the center of the reconstructions we have a high similarity between the reconstruction of the composed phantom and the sum of the separate measurements of vessels and kidneys. This emphasizes that the linearity in Eq.~\eqref{eq:LinearityImagEq} holds true, which is a basic prerequisite of the two-step reconstruction method. The reconstruction results of the subtracted data of the vessels highlights that the projected data $\zbh u_\proj$ is a good approximation of $\zbh u_\vessel$. However, the two-step method is capable of reconstructing even lower concentrated kidneys if the optimal data of the part with higher concentration is provided.

\end{document}

%% file: Mathe-Definitionen.tex
\usepackage{bbm}
\newcommand{\IZ}{\mathbbm{Z}}		
\newcommand{\IR}{\mathbbm{R}}		
\newcommand{\IC}{\mathbbm{C}}		
\newcommand{\cP}{\mathcal{P}}		

\renewcommand{\epsilon}{\varepsilon}					
\renewcommand{\theta}{\vartheta}                  	    
\renewcommand{\phi}{\varphi}							

\newcommand{\set}[1]{\left\{ #1 \right\}}				
\newcommand{\abs}[1]{\left\lvert #1 \right\rvert}		
\newcommand{\norm}[1]{\left\lVert #1 \right\rVert}		

\newcommand{\signatur}[3]{#1:#2\,\rightarrow\,#3}     

\newcommand{\high}{\textup{high}}
\newcommand{\low}{\textup{low}}
\newcommand{\meas}{\textup{meas}}
\newcommand{\thresh}{\textup{thresh}}
\newcommand{\pre}{\textup{pre}}
\newcommand{\proj}{\textup{proj}}
\newcommand{\corr}{\textup{corr}}
\newcommand{\post}{\textup{post}}
\newcommand{\final}{\textup{final}}

\newcommand{\signal}{\textup{signal}}
\newcommand{\artifact}{\textup{artifact}}

\newcommand{\vessel}{\textup{vessel}}

\newcommand{\bet}{ {\boldsymbol\eta} }
\newcommand{\bPhi}{ {\boldsymbol\Phi} }

\newcommand{\zb}[1]{\mbox{\boldmath{${#1}$}}}
\newcommand{\zbs}[1]{\mbox{\boldmath\scriptsize{${#1}$}}}

\renewcommand{\d}{\, \text{d}}

\newcommand{\zbh}[1]{\mbox{$\hat{\zb{#1}}$}} 

\usepackage[binary-units = true,exponent-product = \cdot]{siunitx}
\DeclareSIUnit{\mup}{\text{$\mu_0$}}
\DeclareSIUnit{\sample}{S}
\DeclareSIUnit{\proceduredefinedunit}{p.d.u.}

\definecolor{ibilight}{RGB}{193,216,237}
\definecolor{ibidark}{RGB}{0,73,146}	
\definecolor{uke2}{RGB}{170,156,143} 	
\definecolor{uke3}{RGB}{87,87,86}		
\definecolor{ukesec1}{RGB}{255,223,0}	
\definecolor{ukesec2}{RGB}{239,123,5}	
\definecolor{ukesec3}{RGB}{104,195,205}	
\definecolor{ukesec4}{RGB}{138,189,36}	
\definecolor{tuhh}{RGB}{45,198,214}     
\definecolor{ibidarkBG}{RGB}{227,229,242}   
\definecolor{uke2BG}{RGB}{233,228,225} 	    
\definecolor{uke3BG}{RGB}{230,231,232}	    
\definecolor{ukesec1BG}{RGB}{255,243,190}   
\definecolor{ukesec2BG}{RGB}{254,232,212}   
\definecolor{ukesec3BG}{RGB}{222,241,241}   
\definecolor{ukesec4BG}{RGB}{233,243,222}   

%% file: tikz/tikz_Phantoms.tex
\begin{tikzpicture}[node distance=0.05cm and 0.05cm]
\def\size{3.7cm}
\def\yshift{10pt}

\node (DotsGer) {\includegraphics[width=\size,trim=7cm 0cm 12cm 0cm, clip]
                {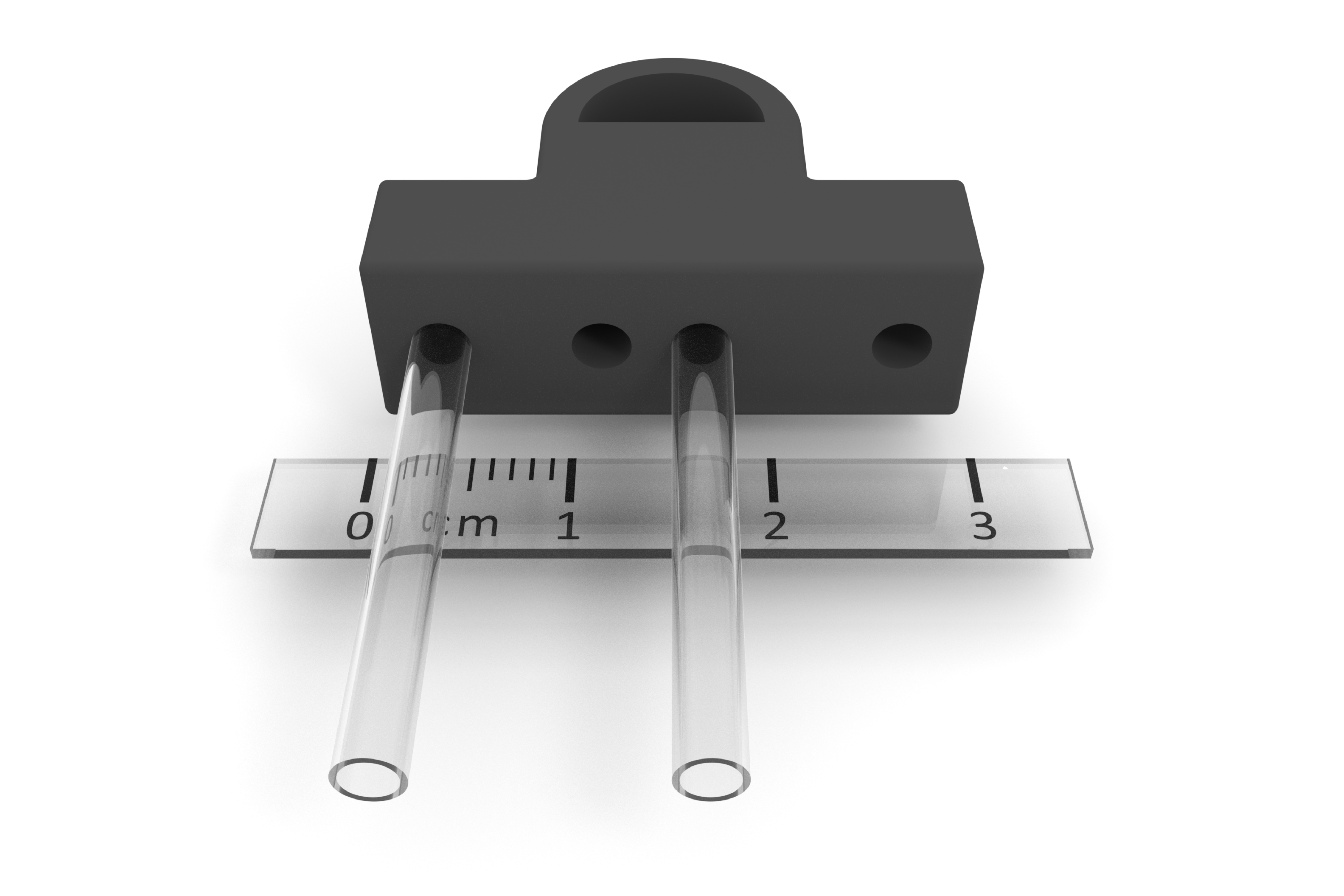}};
\node[right = of DotsGer] (DotsZeich) 
    {\includegraphics[width=\size]{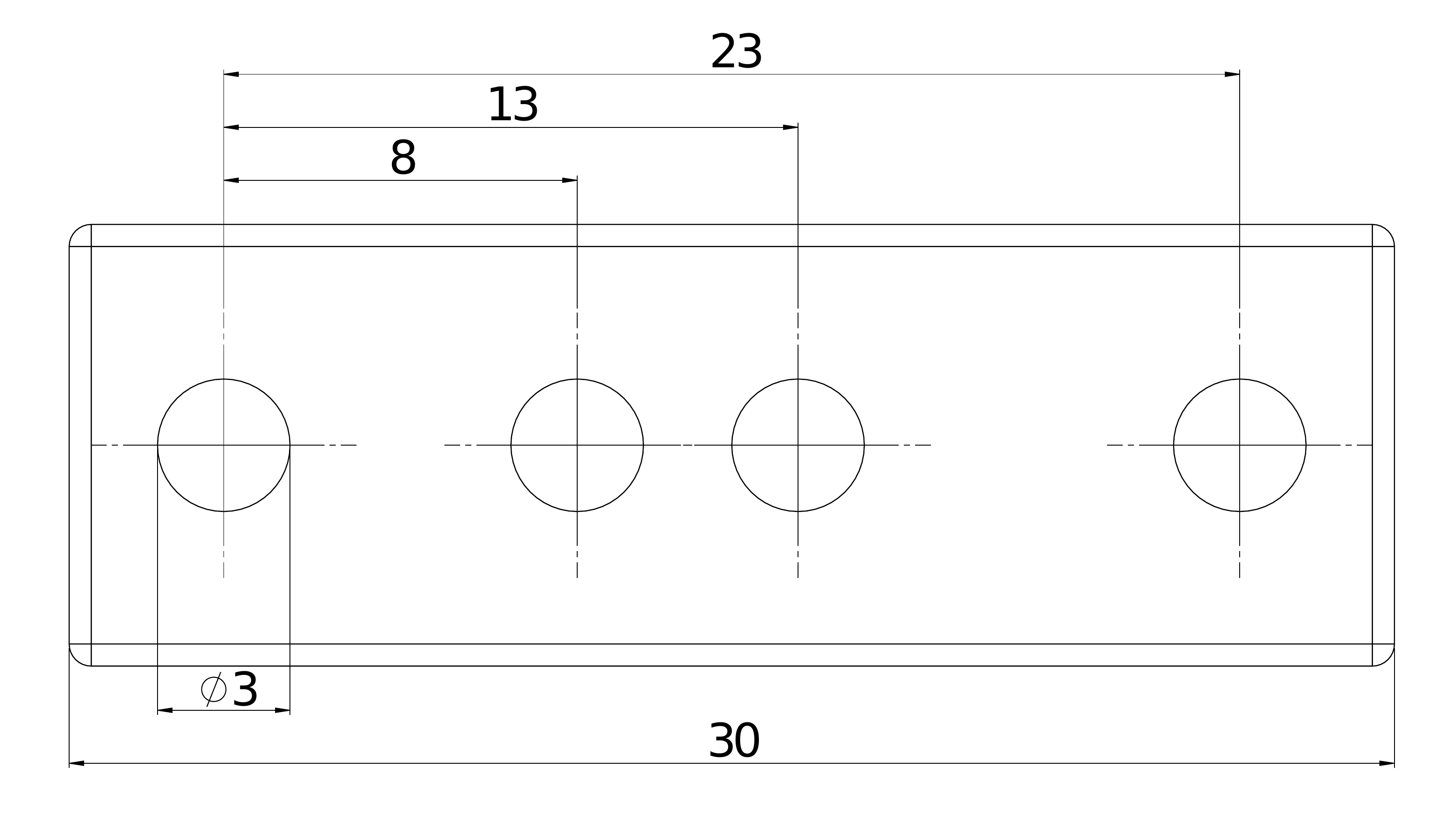}};
\node[right = of DotsZeich] (DotsImg) 
    {\includegraphics[width=\size]{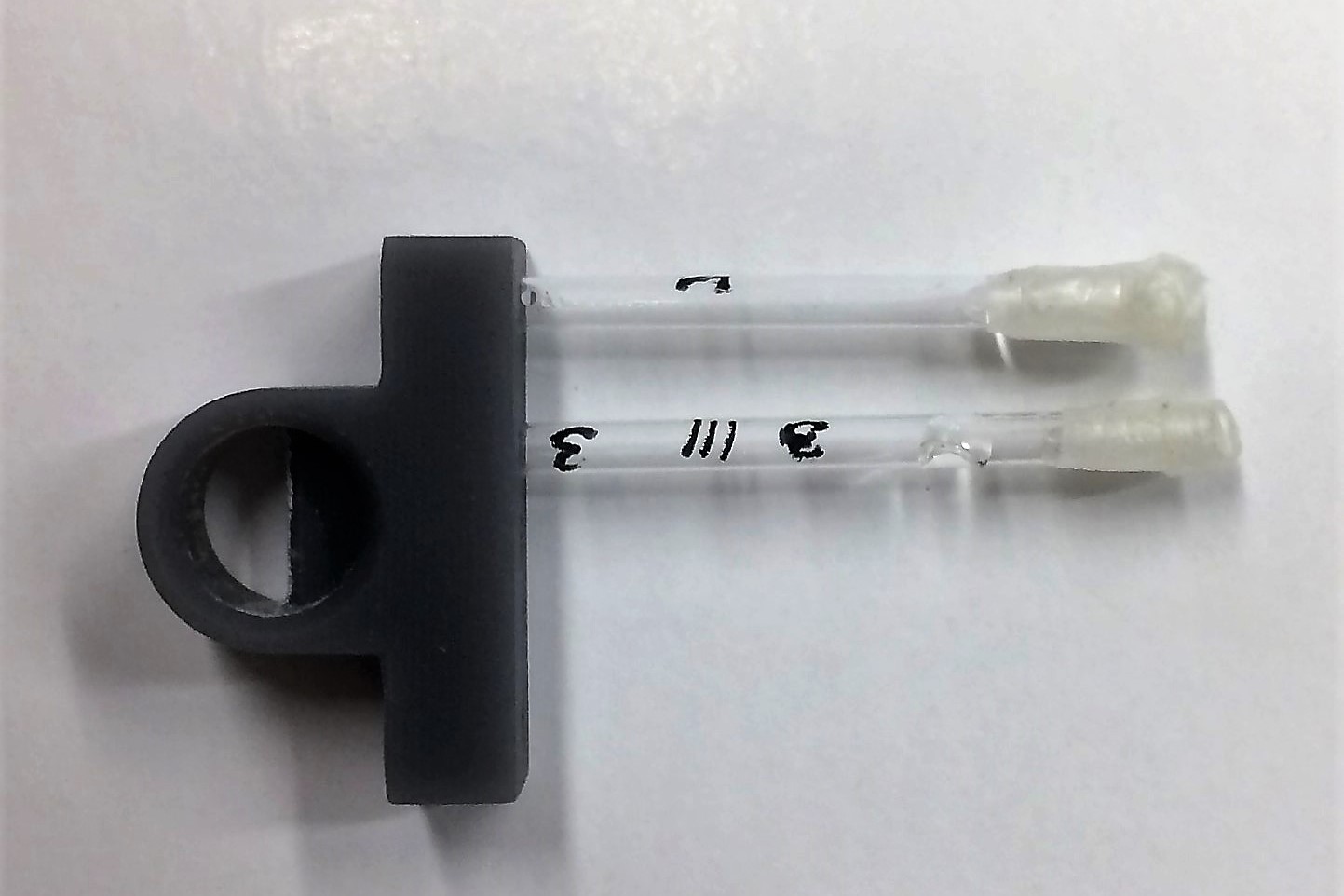}};

\node[below = \yshift of DotsGer] (KidGer)  
    {\includegraphics[width=\size,trim=7cm 0cm 12cm 0cm, clip]{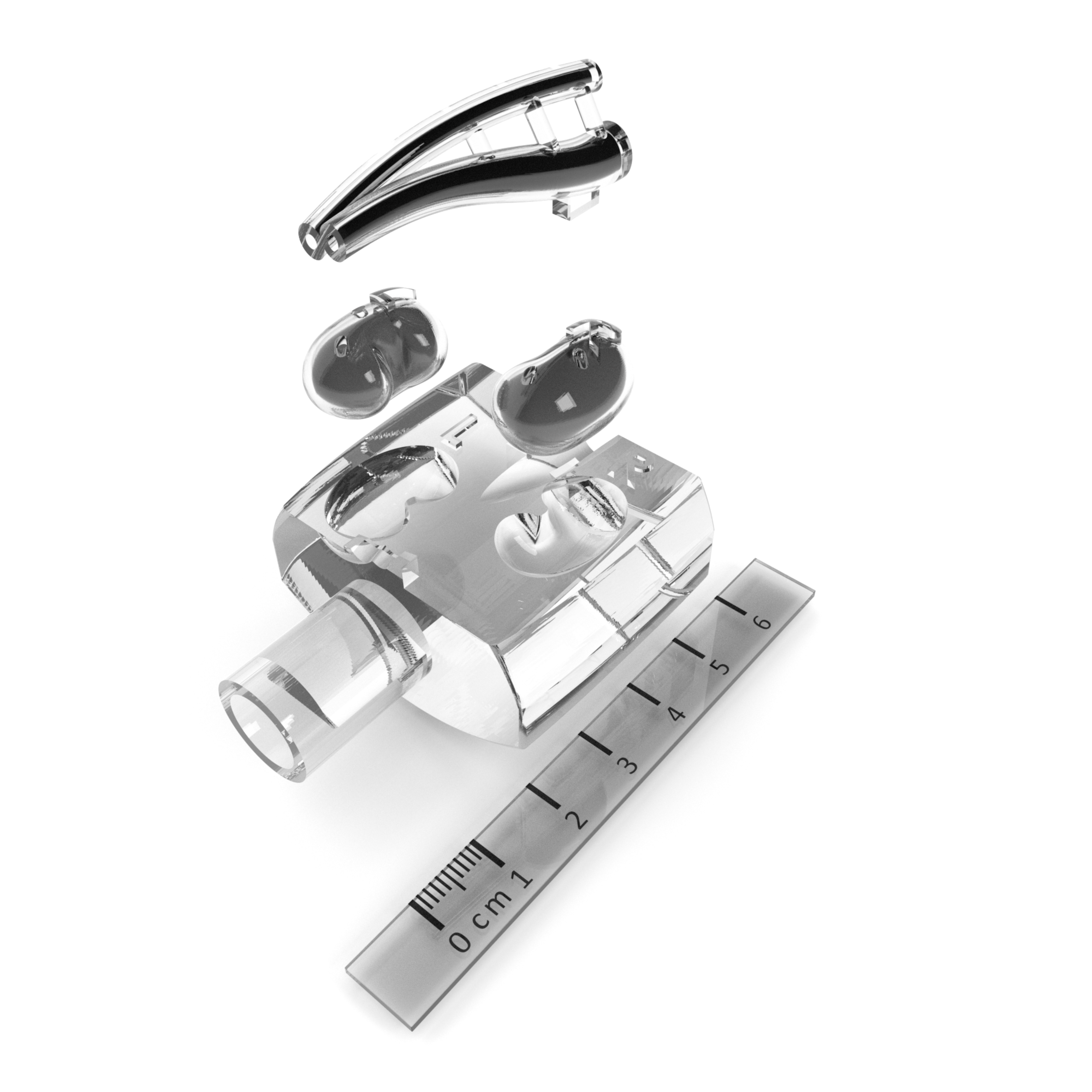}};
\node[right = of KidGer] (KidZeich)  
    {\includegraphics[width=\size,trim=0cm 0cm 0cm 0cm, clip]{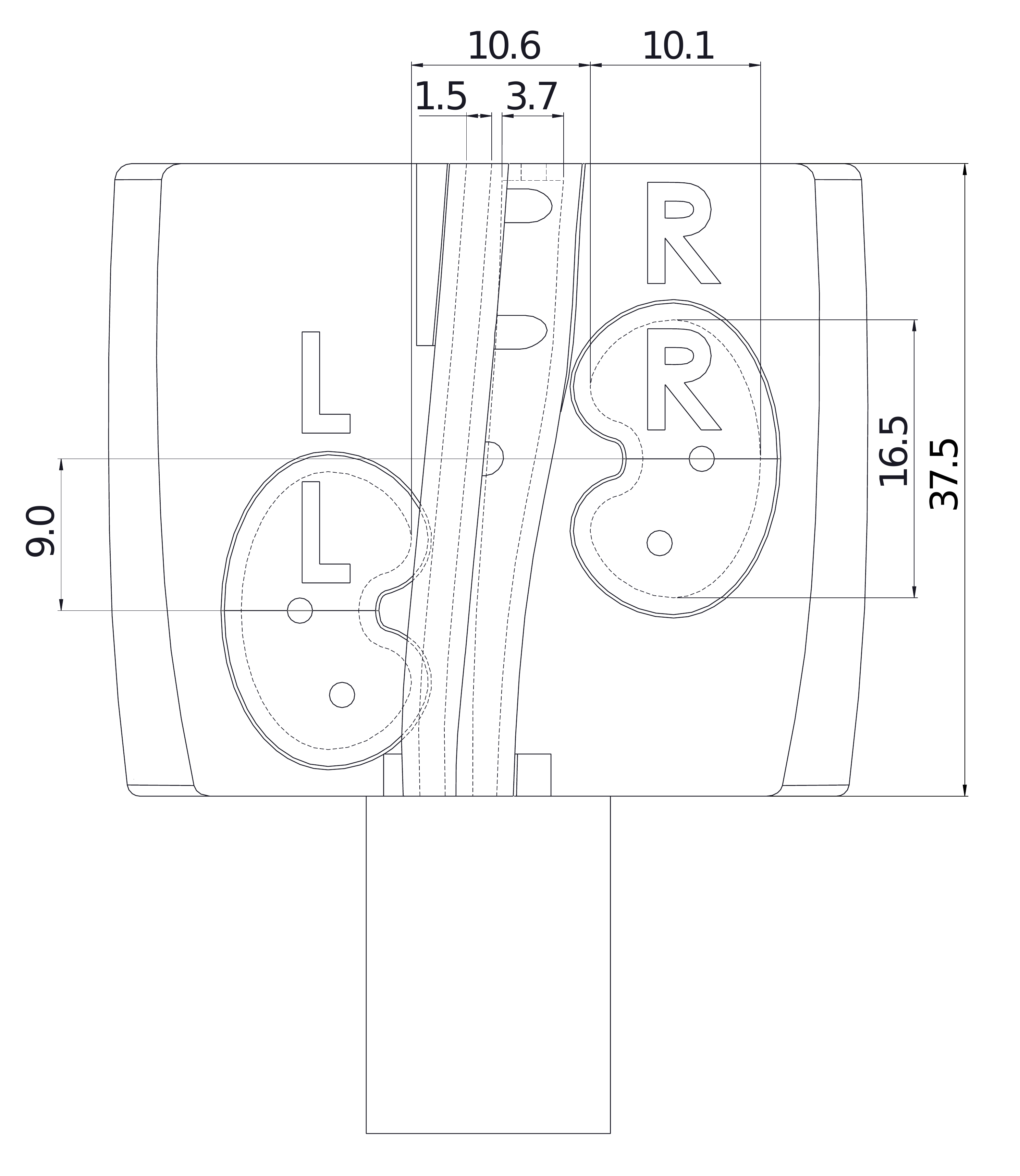}};
\node[right = of KidZeich] (KidImg) 
    {\includegraphics[width=\size]{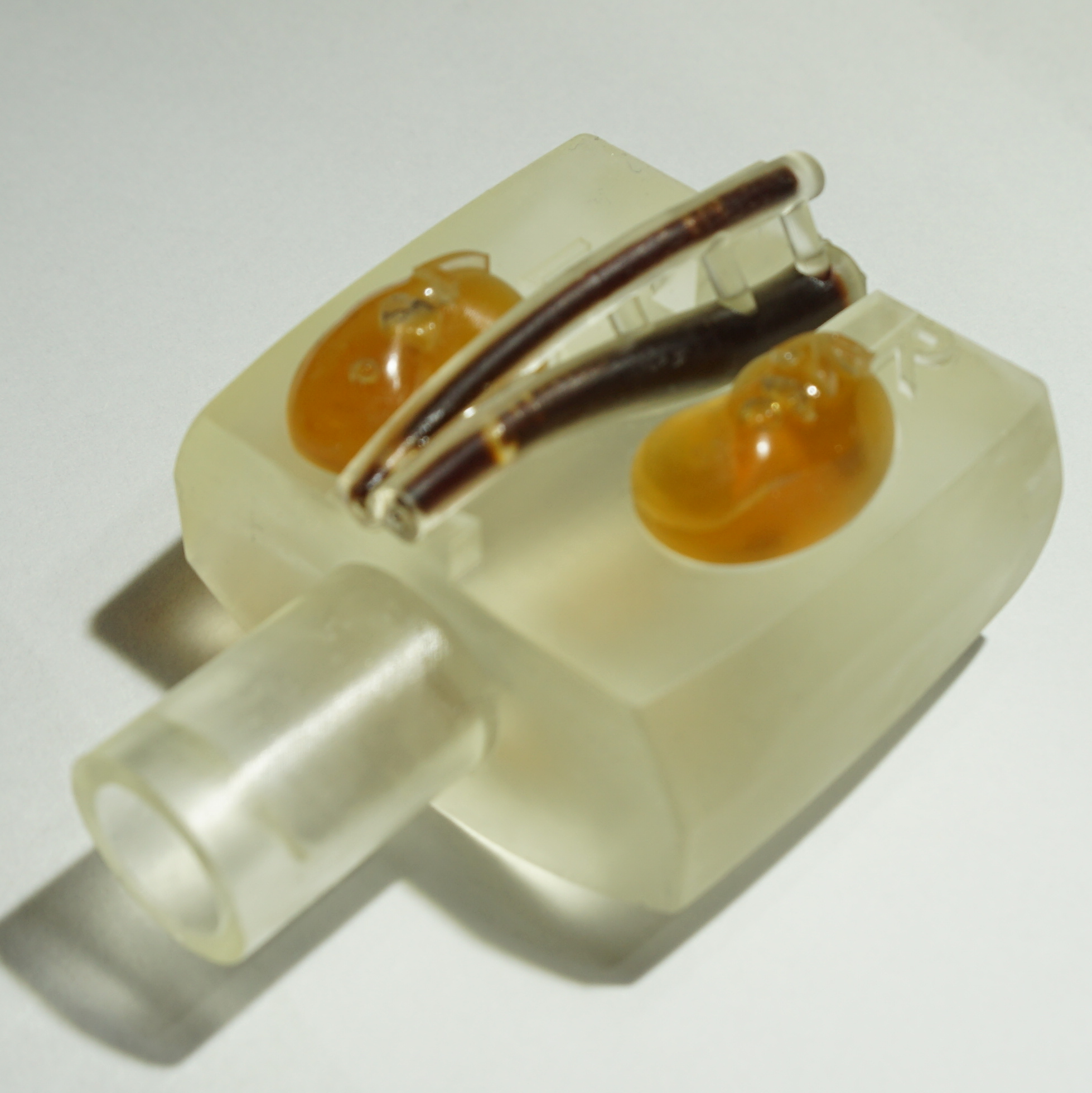}};

\tikzset{
    capt/.style = {rotate=90, text centered, xshift=0.0cm, anchor=south, minimum height=18pt, inner sep=2pt},
}
\node[left = of DotsGer, capt] (DotsCapt) {Dot Phantom};
\node[left = of KidGer, capt] (KidCapt) {Kidney Phantom};

\node[draw=gray,rounded corners, fit=(DotsCapt) (DotsGer) (DotsZeich) (DotsImg)] {};
\node[draw=gray,rounded corners, fit=(KidCapt) (KidGer) (KidZeich) (KidImg)] {};

\end{tikzpicture}

%% file: tikz/tikz_SAR.tex
\begin{tikzpicture}[node distance=-0.05cm and -0.05cm]
\def\sizex{10cm}
\def\sizey{5.5cm}

\pgfplotsset{footnotesize,
             grid=both,
             width=\sizex,
             height=\sizey,
             extra y ticks={1},
             axis background/.style={fill=white},
             ylabel=SAR,
             legend style={at={(1.01,0.5)},anchor=west,
						   cells={anchor=west},}},

\tikzset{every mark/.append style={scale=0.5}}

\begin{axis}[
    xtick={1,2,...,12},
    extra y tick labels={},
    xlabel=Sample number,
    ]
      
    \addplot [ukesec4!70, mark=diamond*,mark options={scale=0.8},
             line width=1.5] table [x = i,y = Single,col sep=comma] {data/SAR_TwoDots.csv};
    \addlegendentry{Single sample};  
    
    \addplot [ukesec2!90!uke3, mark=*,densely dotted,
             line width=1.5] table [x = i,y = Regular_d1,col sep=comma] {data/SAR_TwoDots.csv};
    \addlegendentry{Regular - \SI{5}{\milli\meter}};
    \addplot [ukesec2!90!uke3, mark=*,
             line width=1.5] table [x = i,y = Regular_d2,col sep=comma] {data/SAR_TwoDots.csv};
    \addlegendentry{Regular - \SI{10}{\milli\meter}};
    \addplot [ukesec2!90!uke3, mark=*,dashed,
             line width=1.5] table [x = i,y = Regular_d3,col sep=comma] {data/SAR_TwoDots.csv};
    \addlegendentry{Regular - \SI{20}{\milli\meter}};
    
    \addplot [ibidark, mark=square*,densely dotted,
             line width=1.5] table [x = i,y = TwoStep_d1,col sep=comma] {data/SAR_TwoDots.csv};
    \addlegendentry{Two Step - \SI{5}{\milli\meter}};
    \addplot [ibidark, mark=square*,
             line width=1.5] table [x = i,y = TwoStep_d2,col sep=comma] {data/SAR_TwoDots.csv};
    \addlegendentry{Two Step - \SI{10}{\milli\meter}};
    \addplot [ibidark, mark=square*,dashed,
             line width=1.5] table [x = i,y = TwoStep_d3,col sep=comma] {data/SAR_TwoDots.csv};
    \addlegendentry{Two Step - \SI{20}{\milli\meter}};
    
\end{axis}

\end{tikzpicture}

%% file: tikz/tikz_RecoVessel.tex
\begin{tikzpicture}[node distance=2pt and 2pt]
\def\size{3cm}
\def\sizeZ{12/21 * \size}

\tikzset{
  >={Latex[width=0.03*\size,length=0.03*\size]},}

\tikzset{
  every node/.style={anchor=south west,inner sep=0pt},
  capt/.style = {rotate=90, text centered, text width=\size,xshift=0.0cm,anchor=south, minimum height=18pt},
  capt2/.style = {capt, text width=2.0*\size, minimum height=24pt, yshift=18pt},
}

\node[] (Phantom) {\includegraphics[width=\size]{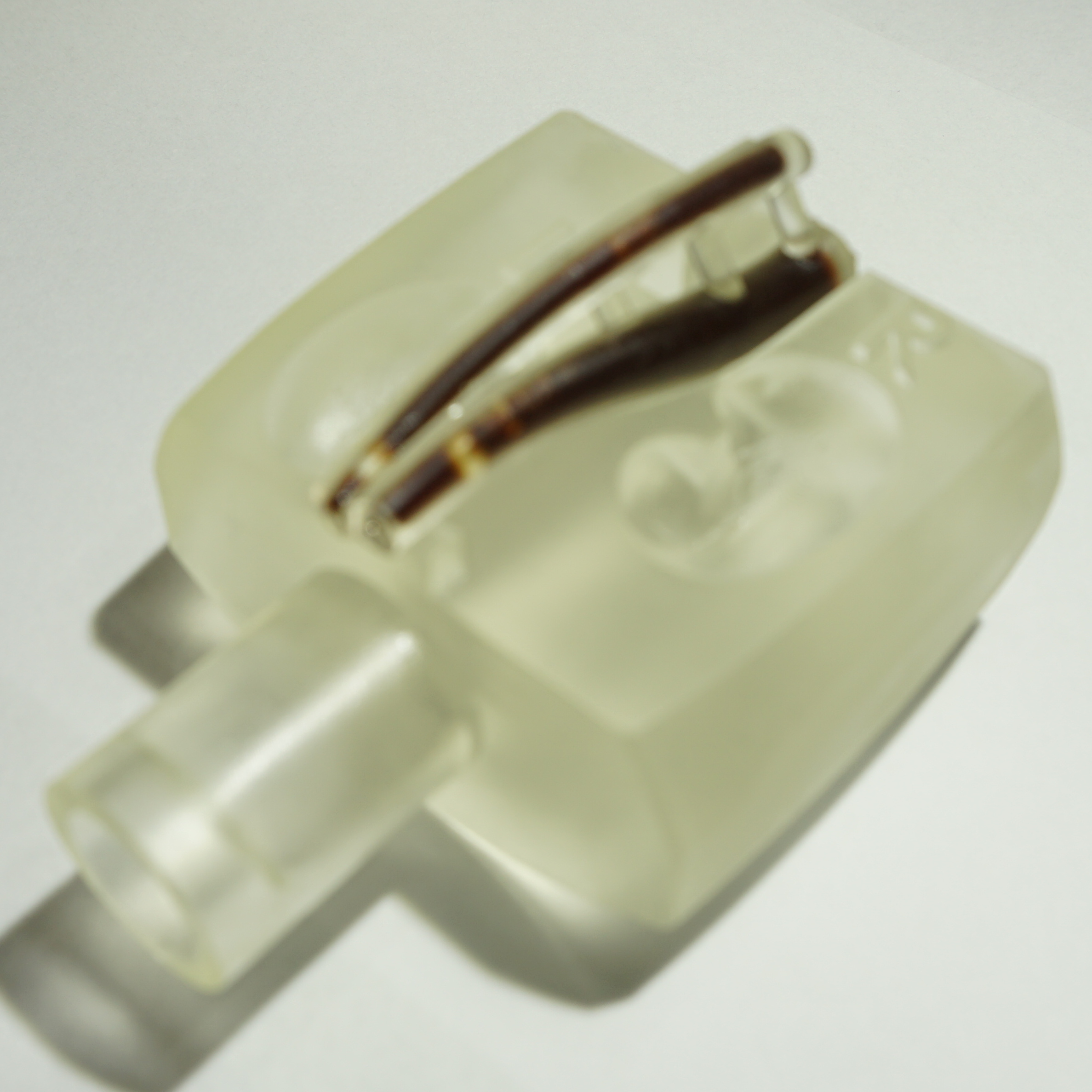}};

\node (RecoXZ) [right = of Phantom]  {\includegraphics[height=\size]{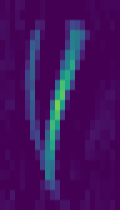}};

\node[right = of RecoXZ] (RecoXY) {\includegraphics[width=\size]{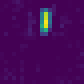}};

\node[above = of Phantom,yshift=2pt] (captPh) {Phantom};
\node[above right = of RecoXZ, xshift=-18pt,yshift=2pt] (captXZ) {Reconstruction};

\coordinate (OPh) at ($(Phantom.south west) +(0.15*\size,0.83*\size)$);
\draw[->] (OPh) -- +(-0.07*\size,-0.07*\size) node[below left] {\footnotesize $x$};
\draw[->] (OPh) -- +(0.0,0.1*\size) node[above=0.5pt] {\footnotesize $z$};
\draw[->] (OPh) -- +(0.07*\size,-0.07*\size) node[below right] {\footnotesize $y$};

\coordinate (OXZ) at ($(RecoXZ.north east) +(-0.05*\size,-0.05*\size)$);
\draw[->,white] (OXZ) -- +(0.0,-0.1*\size) node[below=0.5pt] {\footnotesize $x$};
\draw[->,white] (OXZ) -- +(-0.1*\size,0.0) node[left] {\footnotesize $z$};

\coordinate (OXY) at ($(RecoXY.north west) +(0.05*\size,-0.05*\size)$);
\draw[->,white] (OXY) -- +(0.0,-0.1*\size) node[below=0.5pt] {\footnotesize $x$};
\draw[->,white] (OXY) -- +(0.1*\size,0.0) node[right] {\footnotesize $y$};

\coordinate (LXZ1) at ($(RecoXZ.north east) +(-9.6/24*\sizeZ,0.0)$);
\coordinate (LXZ2) at ($(RecoXZ.north east) +(-9.6/24*\sizeZ,-\size)$);
\draw[white,densely dotted, thick] (LXZ1) -- (LXZ2);

\coordinate (LXY1) at ($(RecoXY.north east) +(-11.05/21*\size,0.0)$);
\coordinate (LXY2) at ($(RecoXY.north east) +(-11.05/21*\size,-\size)$);
\draw[white,densely dotted, thick] (LXY1) -- (LXY2);

\end{tikzpicture}

%% file: main.bbl
\begin{thebibliography}{10}

\bibitem{knopp2017magnetic}
T.~Knopp, N.~Gdaniec, and M.~M{\"o}ddel, ``Magnetic particle imaging: from
  proof of principle to preclinical applications,'' {\em Physics in Medicine \&
  Biology}, vol.~62, no.~14, p.~R124, 2017.

\bibitem{Weizenecker2007PhysMedBio}
J.~Weizenecker, J.~Borgert, and B.~Gleich, ``A simulation study on the
  resolution and sensitivity of magnetic particle imaging,'' {\em Physics in
  Medicine and Biology}, vol.~52, no.~21, pp.~6363 -- 6374, 2007.

\bibitem{vogel2016first}
P.~Vogel, M.~R{\"u}ckert, P.~Klauer, W.~Kullmann, P.~Jakob, and V.~Behr,
  ``First in vivo traveling wave magnetic particle imaging of a beating mouse
  heart,'' {\em Physics in Medicine \& Biology}, vol.~61, no.~18, p.~6620,
  2016.

\bibitem{Graeser2017SR}
M.~Graeser, T.~Knopp, P.~Szwargulski, T.~Friedrich, A.~von Gladiss, M.~Kaul,
  K.~M. Krishnan, H.~Ittrich, G.~Adam, and T.~M. Buzug, ``Towards picogram
  detection of superparamagnetic iron-oxide particles using a gradiometric
  receive coil,'' {\em Scientific Reports}, vol.~7, p.~6872, 2017.

\bibitem{ludewig2017magnetic}
P.~Ludewig, N.~Gdaniec, J.~Sedlacik, N.~D. Forkert, P.~Szwargulski, M.~Graeser,
  G.~Adam, M.~G. Kaul, K.~M. Krishnan, R.~M. Ferguson, {\em et~al.}, ``Magnetic
  particle imaging for real-time perfusion imaging in acute stroke,'' {\em ACS
  nano}, vol.~11, no.~10, pp.~10480--10488, 2017.

\bibitem{Szwargulski2020}
P.~Szwargulski, M.~Wilmes, E.~Javidi, F.~Thieben, M.~Graeser, M.~Koch,
  C.~Gruettner, G.~Adam, C.~Gerloff, T.~Magnus, T.~Knopp, and P.~Ludewig,
  ``Monitoring intracranial cerebral hemorrhage using multicontrast real-time
  magnetic particle imaging,'' {\em ACS Nano}, vol.~14, no.~10,
  pp.~13913--13923, 2020.
\newblock PMID: 32941000.

\bibitem{yu2017magneticB}
E.~Y. Yu, P.~Chandrasekharan, R.~Berzon, Z.~W. Tay, X.~Y. Zhou, A.~P. Khandhar,
  R.~M. Ferguson, S.~J. Kemp, B.~Zheng, P.~W. Goodwill, {\em et~al.},
  ``Magnetic particle imaging for highly sensitive, quantitative, and safe in
  vivo gut bleed detection in a murine model,'' {\em ACS nano}, vol.~11,
  no.~12, pp.~12067--12076, 2017.

\bibitem{arami2017tomographic}
H.~Arami, E.~Teeman, A.~Troksa, H.~Bradshaw, K.~Saatchi, A.~Tomitaka, S.~S.
  Gambhir, U.~O. H{\"a}feli, D.~Liggitt, and K.~M. Krishnan, ``Tomographic
  magnetic particle imaging of cancer targeted nanoparticles,'' {\em
  Nanoscale}, vol.~9, no.~47, pp.~18723--18730, 2017.

\bibitem{vaalma2017magnetic}
S.~Vaalma, J.~Rahmer, N.~Panagiotopoulos, R.~L. Duschka, J.~Borgert,
  J.~Barkhausen, F.~M. Vogt, and J.~Haegele, ``{Magnetic particle imaging
  (MPI): Experimental quantification of vascular stenosis using stationary
  stenosis phantoms},'' {\em PloS one}, vol.~12, no.~1, p.~e0168902, 2017.

\bibitem{sedlacik2016magnetic}
J.~Sedlacik, A.~Fr{\"o}lich, J.~Spallek, N.~D. Forkert, T.~D. Faizy, F.~Werner,
  T.~Knopp, D.~Krause, J.~Fiehler, and J.-H. Buhk, ``Magnetic particle imaging
  for high temporal resolution assessment of aneurysm hemodynamics,'' {\em PloS
  one}, vol.~11, no.~8, p.~e0160097, 2016.

\bibitem{kaul2018magnetic}
M.~G. Kaul, J.~Salamon, T.~Knopp, H.~Ittrich, G.~Adam, H.~Weller, and C.~Jung,
  ``Magnetic particle imaging for in vivo blood flow velocity measurements in
  mice,'' {\em Physics in Medicine \& Biology}, vol.~63, no.~6, p.~064001,
  2018.

\bibitem{Vogel2020}
P.~{Vogel}, M.~A. {Rückert}, T.~{Kampf}, S.~{Herz}, A.~{Stang}, L.~{Wöckel},
  T.~A. {Bley}, S.~{Dutz}, and V.~C. {Behr}, ``Superspeed bolus visualization
  for vascular magnetic particle imaging,'' {\em IEEE Transactions on Medical
  Imaging}, vol.~39, no.~6, pp.~2133--2139, 2020.

\bibitem{zhou2017first}
X.~Y. Zhou, K.~E. Jeffris, Y.~Y. Elaine, B.~Zheng, P.~W. Goodwill, P.~Nahid,
  and S.~M. Conolly, ``First in vivo magnetic particle imaging of lung
  perfusion in rats,'' {\em Physics in Medicine \& Biology}, vol.~62, no.~9,
  p.~3510, 2017.

\bibitem{bulte2015quantitative}
J.~W. Bulte, P.~Walczak, M.~Janowski, K.~M. Krishnan, H.~Arami, A.~Halkola,
  B.~Gleich, and J.~Rahmer, ``{Quantitative Hot Spot Imaging of Transplanted
  Stem Cells using Superparamagnetic Tracers and Magnetic Particle Imaging
  (MPI)},'' {\em Tomography: a journal for imaging research}, vol.~1, no.~2,
  p.~91, 2015.

\bibitem{cooley2018rodent}
C.~Z. Cooley, J.~B. Mandeville, E.~E. Mason, E.~T. Mandeville, and L.~L. Wald,
  ``{Rodent Cerebral Blood Volume (CBV) changes during hypercapnia observed
  using Magnetic Particle Imaging (MPI) detection},'' {\em NeuroImage}, 2018.

\bibitem{haegele2016magnetic}
J.~Haegele, N.~Panagiotopoulos, S.~Cremers, J.~Rahmer, J.~Franke, R.~L.
  Duschka, S.~Vaalma, M.~Heidenreich, J.~Borgert, P.~Borm, {\em et~al.},
  ``Magnetic particle imaging: A resovist based marking technology for guide
  wires and catheters for vascular interventions,'' {\em IEEE transactions on
  medical imaging}, vol.~35, no.~10, pp.~2312--2318, 2016.

\bibitem{salamon2016magnetic}
J.~Salamon, M.~Hofmann, C.~Jung, M.~G. Kaul, F.~Werner, K.~Them, R.~Reimer,
  P.~Nielsen, A.~vom Scheidt, G.~Adam, {\em et~al.}, ``{Magnetic
  particle/magnetic resonance imaging: In-vitro MPI-guided real time catheter
  tracking and 4D angioplasty using a road map and blood pool tracer
  approach},'' {\em PloS one}, vol.~11, no.~6, p.~e0156899, 2016.

\bibitem{herz2018magnetic}
S.~Herz, P.~Vogel, P.~Dietrich, T.~Kampf, M.~A. R{\"u}ckert, R.~Kickuth, V.~C.
  Behr, and T.~A. Bley, ``Magnetic particle imaging guided real-time
  percutaneous transluminal angioplasty in a phantom model,'' {\em
  Cardiovascular and interventional radiology}, pp.~1--6, 2018.

\bibitem{haegele2016multi}
J.~Haegele, S.~Vaalma, N.~Panagiotopoulos, J.~Barkhausen, F.~M. Vogt,
  J.~Borgert, and J.~Rahmer, ``Multi-color magnetic particle imaging for
  cardiovascular interventions,'' {\em Physics in Medicine and Biology},
  vol.~61, no.~16, p.~N415, 2016.

\bibitem{loewa2016concentration}
N.~L\"owa, P.~Radon, O.~Kosch, and F.~Wiekhorst, ``{Concentration dependent MPI
  tracer performance},'' {\em International Journal on Magnetic Particle
  Imaging}, vol.~2, no.~1, 2016.

\bibitem{erb2018mathematical}
W.~Erb, A.~Weinmann, M.~Ahlborg, C.~Brandt, G.~Bringout, T.~M. Buzug,
  J.~Frikel, C.~Kaethner, T.~Knopp, T.~M{\"a}rz, {\em et~al.}, ``Mathematical
  analysis of the 1d model and reconstruction schemes for magnetic particle
  imaging,'' {\em Inverse Problems}, vol.~34, no.~5, p.~055012, 2018.

\bibitem{Knopp2010PhysMedBio}
T.~Knopp, J.~Rahmer, T.~F. Sattel, S.~Biederer, J.~Weizenecker, B.~Gleich,
  J.~Borgert, and T.~M. Buzug, ``Weighted iterative reconstruction for magnetic
  particle imaging,'' {\em Physics in Medicine and Biology}, vol.~55, no.~6,
  pp.~1577 -- 1589, 2010.

\bibitem{FrankeHybridMRMPI2016}
J.~Franke, U.~Heinen, H.~Lehr, A.~Weber, F.~Jaspard, W.~Ruhm, M.~Heidenreich,
  and V.~Schulz, ``System characterization of a highly integrated preclinical
  hybrid {MPI}-{MRI} scanner,'' {\em IEEE Transactions on Medical Imaging},
  vol.~35, no.~9, pp.~1993--2004, 2016.

\bibitem{Herz2017}
S.~Herz, P.~Vogel, T.~Kampf, M.~A. R{\"u}ckert, V.~C. Behr, and T.~A. Bley,
  ``{Selective signal suppression in traveling wave MPI: Focusing on areas with
  low concentration of magnetic particles},'' {\em International Journal on
  Magnetic Particle Imaging}, vol.~3, no.~2, 2017.

\bibitem{Oeff1955}
K.~Oeff and A.~K{\"o}nig, ``{Das Blutvolumen einiger Rattenorgane und ihre
  Restblutmenge nach Entbluten bzw. Durchsp{\"u}lung. Bestimmung mit
  P32-markierten Erythrocyten},'' {\em Naunyn-Schmiedebergs Archiv f{\"u}r
  experimentelle Pathologie und Pharmakologie}, vol.~226, pp.~98--102, Jan
  1955.

\bibitem{graeser2013analog}
M.~Graeser, T.~Knopp, M.~Gr{\"u}ttner, T.~F. Sattel, and T.~M. Buzug, ``Analog
  receive signal processing for magnetic particle imaging,'' {\em Medical
  physics}, vol.~40, no.~4, 2013.

\bibitem{Kluth2019}
T.~Kluth, P.~Szwargulski, and T.~Knopp, ``Towards accurate modeling of the
  multidimensional magnetic particle imaging physics,'' {\em New Journal of
  Physics}, vol.~21, p.~103032, oct 2019.

\bibitem{Rahmer2012TMI}
J.~Rahmer, J.~Weizenecker, B.~Gleich, and J.~Borgert, ``Analysis of a {3-D}
  system function measured for magnetic particle imaging,'' {\em IEEE
  Transactions on Medical Imaging}, vol.~31, no.~6, pp.~1289 -- 1299, 2012.

\bibitem{straub2018joint}
M.~Straub and V.~Schulz, ``Joint reconstruction of tracer distribution and
  background in magnetic particle imaging,'' {\em IEEE transactions on medical
  imaging}, vol.~37, no.~5, pp.~1192--1203, 2018.

\bibitem{paysen2020characterization}
H.~Paysen, O.~Kosch, J.~Wells, N.~Loewa, and F.~Wiekhorst, ``Characterization
  of noise and background signals in a magnetic particle imaging system,'' {\em
  Physics in Medicine \& Biology}, 2020.

\bibitem{szwargulski2017influence}
P.~Szwargulski and T.~Knopp, ``Influence of the receive channel number on the
  spatial resolution in magnetic particle imaging,'' {\em International Journal
  on Magnetic Particle Imaging}, vol.~3, no.~1, 2017.

\bibitem{them2016sensitivity}
K.~Them, M.~G. Kaul, C.~Jung, M.~Hofmann, T.~Mummert, F.~Werner, and T.~Knopp,
  ``Sensitivity enhancement in magnetic particle imaging by background
  subtraction,'' {\em IEEE transactions on medical imaging}, vol.~35, no.~3,
  pp.~893--900, 2016.

\bibitem{knopp2016online}
T.~Knopp and M.~Hofmann, ``Online reconstruction of 3d magnetic particle
  imaging data,'' {\em Physics in Medicine \& Biology}, vol.~61, no.~11,
  p.~N257, 2016.

\bibitem{vogel2017low}
P.~Vogel, S.~Herz, T.~Kampf, M.~A. R{\"u}ckert, T.~A. Bley, and V.~C. Behr,
  ``Low latency real-time reconstruction for {MPI} systems,'' {\em
  International Journal on Magnetic Particle Imaging}, vol.~3, no.~2, 2017.

\bibitem{fleming1990}
H.~E. Fleming, ``Equivalence of regularization and truncated iteration in the
  solution of ill-posed image reconstruction problems,'' {\em Linear Algebra
  and its Applications}, vol.~130, pp.~133 -- 150, 1990.

\bibitem{Knopp2019BG}
T.~Knopp, N.~Gdaniec, R.~Rehr, M.~Graeser, and T.~Gerkmann, ``Correction of
  linear system drifts in magnetic particle imaging,'' {\em Physics in Medicine
  {\&} Biology}, vol.~64, p.~125013, jun 2019.

\bibitem{Exner2019}
M.~Exner, P.~Szwargulski, T.~Knopp, M.~Graeser, and P.~Ludewig, ``3d printed
  anatomical model of a rat for medical imaging,'' {\em Current Directions in
  Biomedical Engineering}, vol.~5, no.~1, pp.~187 -- 190, 2019.

\bibitem{Kaczmarz1937}
S.~Kaczmarz, ``Angen\"aherte {A}ufl\"osung von {S}ystemen linearer
  {G}leichungen,'' {\em Bulletin of the International Academy Polonica Sciences
  Letters A}, vol.~35, pp.~355 -- 357, 1937.

\bibitem{bezanson2017julia}
J.~Bezanson, A.~Edelman, S.~Karpinski, and V.~B. Shah, ``Julia: A fresh
  approach to numerical computing,'' {\em SIAM review}, vol.~59, no.~1,
  pp.~65--98, 2017.

\bibitem{knopp2019mpireco}
T.~Knopp, P.~Szwargulski, F.~Griese, M.~Grosser, M.~Boberg, and M.~M{\"o}ddel,
  ``{MPIReco.jl}: Julia package for image reconstruction in {MPI},'' {\em
  International Journal on Magnetic Particle Imaging}, vol.~4, no.~2, 2019.

\bibitem{moreland2009diverging}
K.~Moreland, ``Diverging color maps for scientific visualization,'' in {\em
  International Symposium on Visual Computing}, pp.~92--103, Springer, 2009.

\bibitem{Knopp2015PhysMedBiol}
T.~Knopp, K.~Them, M.~Kaul, and N.~Gdaniec, ``Joint reconstruction of
  non-overlapping magnetic particle imaging focus-field data,'' {\em Physics in
  Medicine and Biology}, vol.~60, p.~L15, 2015.

\bibitem{gdaniec2017fast}
N.~Gdaniec, P.~Szwargulski, and T.~Knopp, ``Fast multiresolution data
  acquisition for magnetic particle imaging using adaptive feature detection,''
  {\em Medical physics}, vol.~44, no.~12, pp.~6456--6460, 2017.

\bibitem{weber2015artifact}
A.~Weber, F.~Werner, J.~Weizenecker, T.~Buzug, and T.~Knopp, ``Artifact free
  reconstruction with the system matrix approach by overscanning the
  field-free-point trajectory in magnetic particle imaging,'' {\em Physics in
  Medicine \& Biology}, vol.~61, no.~2, p.~475, 2015.

\bibitem{ahlborg2016using}
M.~Ahlborg, C.~Kaethner, T.~Knopp, P.~Szwargulski, and T.~Buzug, ``Using data
  redundancy gained by patch overlaps to reduce truncation artifacts in
  magnetic particle imaging,'' {\em Physics in Medicine and Biology}, vol.~61,
  no.~12, pp.~4583--4598, 2016.

\bibitem{knopp2008trajectory}
T.~Knopp, S.~Biederer, T.~Sattel, J.~Weizenecker, B.~Gleich, J.~Borgert, and
  T.~Buzug, ``Trajectory analysis for magnetic particle imaging,'' {\em Physics
  in Medicine \& Biology}, vol.~54, no.~2, p.~385, 2008.

\bibitem{werner2017first}
F.~Werner, N.~Gdaniec, and T.~Knopp, ``First experimental comparison between
  the cartesian and the lissajous trajectory for magnetic particle imaging,''
  {\em Physics in Medicine \& Biology}, vol.~62, no.~9, p.~3407, 2017.

\bibitem{Storath2017}
M.~{Storath}, C.~{Brandt}, M.~{Hofmann}, T.~{Knopp}, J.~{Salamon}, A.~{Weber},
  and A.~{Weinmann}, ``Edge preserving and noise reducing reconstruction for
  magnetic particle imaging,'' {\em IEEE Transactions on Medical Imaging},
  vol.~36, no.~1, pp.~74--85, 2017.

\end{thebibliography}
